\documentclass[useAMS]{mn2e}
\usepackage{latexsym,graphicx}

\DeclareGraphicsExtensions{.pdf,.png,.jpg}

\usepackage{color}
\usepackage{amsmath}
\usepackage{bm}

\newcommand\msun{\rm\,M_\odot}

\newcommand\lsun{\rm\,L_\odot}

\DeclareMathOperator{\sgn}{sgn}

\title[Dust arcs around the stars]{Infrared dust arcs around the stars: I. effect of the radiation pressure}

\author[O. A.~Katushkina et al.]{O. A.~Katushkina$^{1}$\thanks{E-mail: okat@iki.rssi.ru},
  V. V.~Izmodenov$^{1,2,3}$ \\
$^{1}$Space Research Institute of Russian Academy of Sciences, Profsoyuznaya Str. 84/32, Moscow, 117335, Russia\\
$^{2}$Lomonosov Moscow State University, GSP-1, Leninskie Gory, Moscow, 119991, Russia\\
$^{3}$Institute for Problems in Mechanics, prosp. Vernadskogo 101, block 1, Moscow, 119526, Russia\\
}
\begin{document}

\def\vecv{\mbox{$\textbf{v}_p$}}
\def\vecV{\mbox{$\textbf{V}$}}
\def\vecB{\mbox{$\textbf{B}$}}
\def\vecq{\mbox{$\textbf{q}$}}
\def\vecQ{\mbox{$\textbf{Q}$}}

\date{Accepted 2019 April 16. Received 2019 April 5; in original form 2019 January 23}

\pagerange{\pageref{firstpage}--\pageref{lastpage}} \pubyear{2019}

\maketitle

\label{firstpage}

\begin{abstract}
In this paper we consider the distribution of the interstellar dust in the vicinity of the star under an influence of the stellar gravitation and radiation pressure. This study is applicable to the stars with relatively weak stellar wind and strong radiation, when the stellar radiation swept out the interstellar dust much further from the star than the position of the bow shock created by an interaction of the stellar and interstellar plasma flows. In this case the number density of dust for a certain dust grains radius can be calculated analytically based on the classical ``cold model''. The dust density distribution for the mixture of dust grains with different radii is calculated. We also calculated intensity maps of the thermal infrared emission at 24~$\mu$m from dust due to heating by the stellar radiation. It is shown that the obtained maps of the infrared emission extremely depends on the model parameters: material of dust grains, dust size distribution assumed in the interstellar medium, the approach used to calculate the dust temperature. The bright distinct arc at the intensity maps is seen for graphite dust grains and almost disappears for silicates.  Absolute values of intensity in the case of graphite are several order of magnitudes larger than for silicates due to more intensive heating of graphite. Possible application of the presented theory is proposed as an algorithm for analysis of the observational images of the infrared arc around the star.
\end{abstract}

\begin{keywords} (ISM:)dust, extinction -- (stars:) circumstellar matter -- radiation: dynamics.
\end{keywords}

\section{Introduction}
\label{sec:intro}


Recently there has been a rapid growth of interest in the study of astrospheres -- the regions of interaction between the stellar wind and
the surrounded interstellar medium (ISM). This is due to availability of new observations made by the \emph{Spitzer Space Telescope}, \emph{Wide-field Infrared Survey Explorer (WISE)} and \emph{Herschel Space Observatory} (Peri et al. 2012; Cox et al. 2012;
Kobulnicky et al. 2016).
The most famous example of the astrosphere is the heliosphere around the Sun, that is studied for decades by both
theoretical and observational methods (see, e.g., a recent review by Izmodenov, 2018). New observations of the astrospheres around other stars
with variety of the stellar and interstellar parameters provide a ``zoo'' of different shapes and structures of the astrospheres (Cox et al., 2012).
Therefore, it becomes possible to expand our knowledge about the heliosphere to different configurations and physical conditions.
It could be useful for studying the evolution of the stars and the stellar winds as well as can help to construct the ISM
parameters at different parts of the interstellar medium not only at the vicinity of the Sun.

However, while the heliosphere is available for direct in-situ measurements of the plasma parameters (e.g. by Voyagers spacecraft, see Stone et al., 2013; Burlaga \& Ness, 2014; Burlaga et al., 2018), most
astrospheres are observed in infrared thermal radiation coming from the interstellar dust.
The interstellar dust interacts with the plasma component (through elastic and Coulomb collisions), stellar radiation and magnetic fields that results in non-uniform
dust distribution around the star. Namely, many infrared images of the astrospheres show an existence of arc-like structures around the star (Cox et al., 2012).
At the same time a specific gasdynamic structure of the shock layer is formed in a gas (plasma) component due to an interaction
of the stellar wind flow with the surrounded interstellar matter.

\begin{figure}
\includegraphics[scale=0.6]{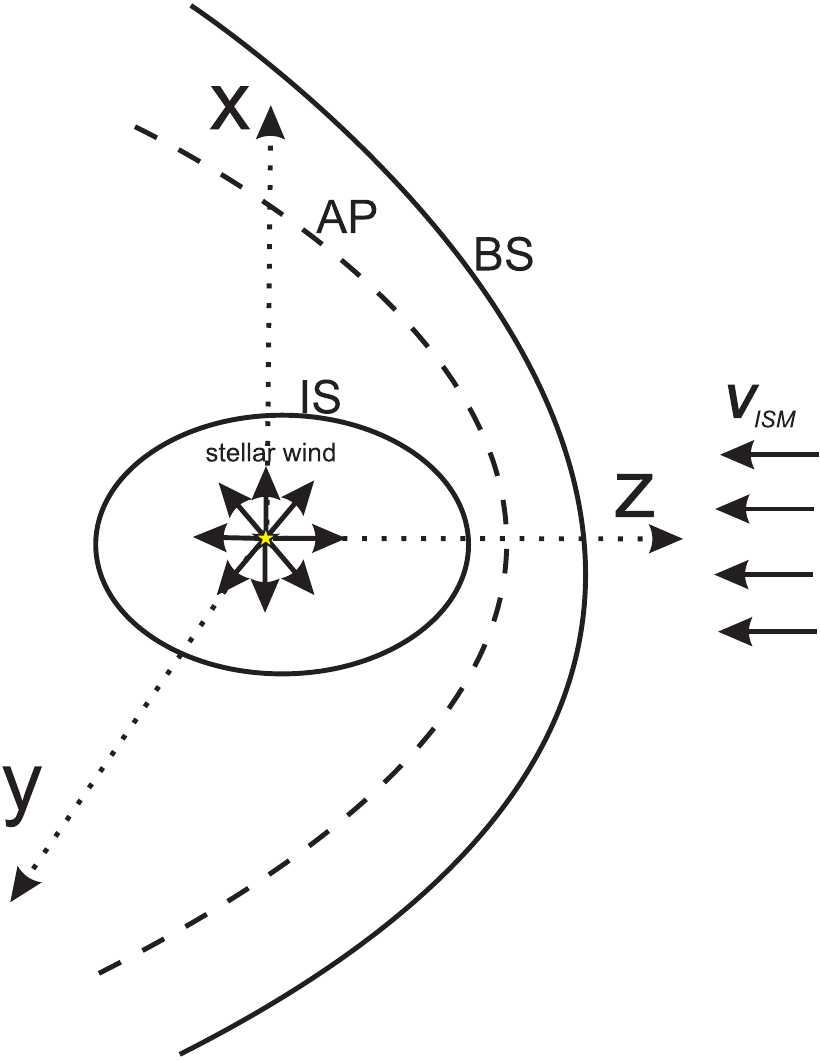}
\centering
\caption{Schematic picture of the astrosphere in the case of the supersonic stellar and interstellar winds.
Surfaces of discontinuities are shown: IS is the inner shock, AP is the astropause (contact or tangential discontinuity), BS is the bow shock.
The coordinate system is marked. $Z$-axis is opposite to the interstellar wind velocity vector ($\textbf{V}_{ISM}$).}
\label{fig:interface}
\end{figure}

Theoretical models of the stellar/solar wind interaction with the surrounding interstellar gas
has been developed since pioneering works by Parker (1961) and Baranov et al. (1970). First models of the temporal evolution of the interstellar bubbles
have been proposed by Castor et al. (1975), Falle (1975), Weaver et al. (1977), McCray (1983).
The typical interaction region of two colliding plasma flows consists of
two shocks (in the case of supersonic character of both stellar and interstellar winds) and contact discontinuity that is called
the astropause (see Fig.~\ref{fig:interface}). The radiative cooling may lead to collapse of the outer shock layer, when a distance
between the bow shock and the astropause is reduced dramatically. In this case the whole structure of the interaction region is
 commonly called as the bow shock, and its shape can be described in the frame of the thin layer approximation (Baranov, Krasnobaev \& Kulikovskii, 1970;
  Wilkin, 1996). However, the interstellar magnetic field may prevent the collapsing of the interaction region (Katushkina et al., 2018; Gvaramadze et al., 2018). Therefore in the presence of a quite strong interstellar magnetic field the thickness of the outer shock layer is not negligible.

The dust distribution in the astrosphere does not necessary follow the gas, therefore an interpretation of the observational infrared images of the thermal dust
 emission from the astrospheres is a puzzling task.

The interstellar dust distribution around the star is determined by the following physical processes:
\begin{itemize}
  \item interaction with the star through the stellar
gravitational attraction and radiation pressure forces (G\'{a}sp\'{a}r et al., 2008);
  \item interaction with the interstellar and stellar magnetic fields though the electromagnetic force (Katushkina et al., 2017, Gvaramadze et al. 2018);
  \item interaction with the gas component (both stellar and interstellar winds) though the collisional and Coulomb drag forces (Baines et al., 1965; Draine \& Salpeter, 1979a);
  \item recharging of dust grains by impinging of protons and electrons, photoelectric emission of stellar and interstellar radiation and
secondary electron emission (Hor\'{a}nyi 1996; Kimura \& Mann, 1998);
  \item possible conglutination and distruction (fragmentation, vaporisation) of dust grains (Draine \& Salpeter, 1979b; Tielens et al., 1994).
\end{itemize}
An efficiency of each of the listed processes is determined by the local conditions and depends on many parameters of the star, gas and dust grains.
Sometimes several processes have a major role and other can be neglected. There could be the limiting case, when one process
is much more important than others, or the case, when all processes should be considered simultaneously. The variety of parameters and possible situations
provides a much complexity of interpretation and analysis of the observations especially because of lack of information about
some physical parameters of the system ``star-gas-dust''. For example, if one has an infrared image of the astrosphere with a distinct arc, the question is
whether this arc seen in dust emission coincides with the bow shock or the astropause in a gas component?
The answer is not obvious and depends on many factors. For example, Ochsendorf et al. (2014a) have shown that the gas and dust are
decoupled in the astrosphere around O9.5V/B0.5V system $\sigma$~Ori~AB. The observed arc in this case can be explained by so called ``dust wave''
that is created due to sweeping of dust by the stellar radiation far away from the star. Mackey et al. (2016) have performed a numerical modelling of the astrosphere around
CD-38$^{\circ}$~11636 in RCW 120 and provided synthetic
intensity maps of infrared dust emission under the assumption of coupling between gas and dust. While Ochsendorf et al. (2014b)
have interpreted mid-infrared arc in H~II region RCW 120 in the context of the dust waves where gas and dust are dynamically decoupled by the radiation pressure.

Many works (Akimkin et al., 2015; Pavlyuchenkov et al. 2013; Ochsendorf et al., 2014a,b; Mackey et al., 2015, 2016; Acreman et al., 2016)
do not take into account the magnetic fields and the electromagnetic force in their modelling of the interstellar dust distribution in the astrospheres, while sometimes the magnetic field can be important (van Marle et al., 2014; Alexashov et al., 2016; Katushkina et al., 2017, 2018).
From the other hand, simultaneous consideration of all physical processes prevents the understanding of the results because
different physical processes may compensate each other and it is very difficult to interpret the results correctly.

In order to understand, which physical processes are important for certain astrosphere one needs to analyze
 the dimensionless parameters of the problem.
 At the beginning of this work we consider an equation of dust motion under an influence of four forces: the stellar gravitational attractive force, the stellar radiation
 repulsion force, the collisional drag force and the electromagnetic force. The dimensionless criteria for importance of each forces are provided.
 After that we focus ourselves on two central forces - the stellar gravitational force and the radiation pressure force.
 Both forces are proportional to $\textbf{r}/r^3$ ($r$ is the distance from the star).
 Therefore, if only these forces exist, then the dust distribution can be described in the frame of the ``classical cold model'' (see, e.g., Fahr, 1968; Meier et al.~1977, Lallement et al., 1985; Izmodenov, 2006) that was used before for
 modelling of the interstellar hydrogen distribution in the heliosphere. The cold model describes the distribution of particles
 moving under an influence of the central force and has a cold (with zero temperature) velocity distribution at infinity (in the ISM).
  By using of the cold model we calculate the interstellar dust distribution in the typical astrosphere and provide synthetic maps of the infrared thermal emission from dust for different parameters of the model. It is studied how the observed arc (its shape and position) in the infrared emission depends on the model parameters,
 and which constrains for the system can be obtained from the visible images of the astrospheres.

\section{Mathematical formulation of the problem: the dust motion equation}
\label{sec:math}

Let us consider a typical astrosphere around the star.
 The interstellar plasma parameters are the following: plasma number density, $n_{p,ISM}$, velocity of the ISM motion relative to the star,
 $\textbf{V}_{ISM}$, temperature of the ISM, $T_{ISM}$, and the interstellar magnetic field, $\textbf{B}_{ISM}$.
 Basically, the interstellar plasma is partially ionized, and the interstellar dust grains may interact
  with neutrals, protons and electrons through collisional drag force.
   The interstellar dust grains penetrate to the astrosphere from the surrounding interstellar medium due to relative motion between
 the star and the ISM.
 The following simplified assumptions are made:
 \begin{itemize}
   \item radius of each dust grain $a$ is constant along its trajectory, i.e. we do not consider any distruction or sublimation of dust;
   \item surface potential $U_d$ and charge $q$ ($q=U_d \cdot a$ in CGS units) of each dust grain are constant along its trajectory, i.e. processes of charging are not considered for the simplicity here although it can be important especially in the region occupied by the stellar wind
       (see e.g. Fig.~2 in Alexashov et al., 2016);
   \item the Coulomb drag force between dust particles and plasma is neglected (expression for Coulomb drag can be found, e.g., in Draine \& Salpeter, 1979a; Hutchinson 2006);
   \item there is no stellar magnetic field, only the interstellar magnetic filed is considered.
 \end{itemize}
Under these assumptions the equation of dust grain motion is the following:
\begin{eqnarray}\label{eq:motion}
\frac{d\textbf{v}_d}{dt}=\frac{1}{m_d}\frac{ \sigma_d \langle Q_{rp}\rangle L_s}{4 \pi c_0}\cdot \frac{\textbf{e}_r}{r^2} - M_s G \cdot \frac{\textbf{e}_r}{r^2} + \\
                       +\frac{2 a^2 k_b n_p T_p \sqrt{\pi} \hat{G}_{pen}}{m_d}\cdot \frac{\textbf{v}_{rel}}{|\textbf{v}_{rel}|}
                       + \frac{q}{c_0 m_d}((\textbf{v}_d-\textbf{v}_p) \times \textbf{B}). \nonumber
\end{eqnarray}
Here, the first term in the right hand part is the stellar radiation pressure force $\textbf{F}_{rad}$, the second term is the stellar gravitational force $\textbf{F}_{g}$, the third term is the collisional drag force $\textbf{F}_{drag}$ due
 to an interaction of dust grain with protons, electrons and neutrals, the forth term is the electromagnetic force $\textbf{F}_{L}$; $\textbf{v}_d$ is the velocity of a dust grain, $m_d=4/3 \pi \rho_d a^3$ is the mass of a dust grain, $\rho_d$ is the density of grain's material, $\sigma_d=4\pi a^2$ is the geometrical cross section of spherical dust grain, $\textbf{e}_r=\textbf{r}/r$ is the unity vector in radial direction, $L_s$ is the stellar luminosity, $M_s$ is the stellar mass, $G$ is the gravitational constant,
$c_0$ is the speed of light, $k_b$ is the Boltzmann constant, $n_p$, $\textbf{v}_p$ and $T_p$ are the number density, velocity and temperature of plasma (gas) component (note, that the gas parameters are equal to
the interstellar parameters far away from the star, but they are disturbed after crossing the bow shock),
$\textbf{B}$ is the magnetic field, $\textbf{v}_{rel}=\textbf{v}_p-\textbf{v}_d$ is the relative velocity
between plasma and dust, $\langle Q_{rp}\rangle$ is the flux weighted mean radiation pressure efficiency factor of dust grain that is:
\[
 \langle Q_{rp}(a)\rangle=\frac{\int_{0}^{+\infty} Q_{rp}(a,\lambda) F_{\lambda} d \lambda}{\int_{0}^{+\infty} F_{\lambda} d \lambda},
\]
where $Q_{rp}(a,\lambda)$ is the radiation pressure efficiency factor, which depends on optical properties of dust grains and can be calculated from
the absorption efficiency $Q_{abs}$ and scattering efficiency $Q_{sca}$:
\[
  Q_{rp}(a,\lambda) = Q_{abs}(a,\lambda)+(1-g)\cdot Q_{sca}(a,\lambda)
\]
and $g$ is the scattering asymmetry factor. $Q_{abs}$, $Q_{sca}$ and $g$ can be calculated by Mie theory (Bohren \& Huffman, 1983)
for certain grain's radius and dust material.
Note that in general case $\langle Q_{rp}\rangle$ depends on the grain's radius, but for the large dust grains with $a\geq1~\mu m$ an approach of the geometrical optics is appropriate and $\langle Q_{rp}\rangle \approx1$ as it will be shown below in Section~2.1.
$F_{\lambda}$ is the spectrum of the stellar radiation and it could be represented by the Planck
function with the effective stellar temperature $T_s$, namely:  $F_{\lambda}\approx B(\lambda, T_s)$.

$\hat{G}_{pen}$ is the dimensionless function appearing in the expression of the collision drag force due to an interaction of dust grain with plasma protons, electrons and neutrals, therefore $\hat{G}_{pen}=\hat{G}(s_p)+\hat{G}(s_e)+\hat{G}(s_n)$, where
\[
 s_i=\sqrt{\frac{m_i v_{rel}^2}{2k_b T_p}}, \, i=\{p,e,n\},
\]
the $m_i$ with $i=\{p,e,n\}$ is the mass of proton, electron and neutral atom correspondingly, here it is assumed that the
plasma component is described in the frame of one fluid approach and, therefore, temperatures of protons, electrons and neutrals are the same and equal to the plasma temperature $T_p$.
A full expression for function $\hat{G}$ can be found in Baines et al. (1965). Draine \& Salpeter (1979a) suggested the following approximation
that is accurate within 1~\% for $0<s_i<+\infty$:
\[
 \hat{G}(s_i)\approx \frac{8 s_i}{3\sqrt{\pi}} \sqrt{1+\frac{9\pi}{64} s_i^2}.
\]
Let us emphasize that the motion equation~(\ref{eq:motion}) should be solved for each certain radius of dust grains presented in the ISM.

Both the stellar radiation pressure force and the gravitational force are proportional to $1/r^2$ and counteract each other. Therefore,
 it is convenient to combine these forces and consider the following dimensionless parameter introduced by Burns, Lamy and Soter (1979):
 \begin{equation}\label{beta}
   \beta(a) = \frac{F_{rad}}{F_{g}} = \frac{3 L_s }{16 \pi c_0 \rho_d G M_s} \cdot \frac{\langle Q_{rp}\rangle}{a}.
 \end{equation}
Thus, equation~(\ref{eq:motion}) can be represented in the following form:

\begin{eqnarray}\label{eq:motion_beta}
\frac{d\textbf{v}_d}{dt}=(\beta(a)-1)G M_s \cdot \frac{\textbf{e}_r}{r^2}
                       +\frac{3 k_b n_p T_p}{2 \sqrt{\pi} \rho_d}\frac{1}{a} \hat{G}_{pen} \cdot \textbf{e}_{rel} + \\
                       + \frac{3 U_d}{4\pi c_0 \rho_d } \frac{1}{a^2}((\textbf{v}_d-\textbf{v}_p) \times \textbf{B}), \nonumber
\end{eqnarray}
here, $\textbf{e}_{rel}=\textbf{v}_{rel}/|\textbf{v}_{rel}|$ is the unity vector of relative velocity and we also substitute expressions for $m_d$ and $q$
as functions of grain's radius.

The boundary condition for the dust velocity in the undisturbed ISM is the following:
\[
 \textbf{v}_d=\textbf{V}_{ISM}.
\]
I.e. the velocity distribution function of dust grains in the ISM is the $\delta$-function: $\delta(\textbf{v}_d-\textbf{V}_{ISM})$.

\subsection{Calculation of parameter $\beta(a)$}
\label{sec:betta}

In order to calculate $\beta(a)$ for certain star and dust grain one needs to know: 1) the stellar parameters --
luminosity $L_s$, mass $M_s$, effective temperature $T_s$ and 2) parameters of the dust grains -- radius $a$, density of material $\rho_d$
and optical constants for calculations of the absorption and scattering efficiencies.


In our calculations we take, as an example, parameters of the system $\sigma$~Ori~AB (HD 37468) that is consistent of 3 stars $\sigma$~Ori~Aa, $\sigma$~Ori~Ab and $\sigma$~Ori~B.
Parameters of these stars are presented by Sim\'{o}n-D\'{i}az et al. (2015). The separation between stars is negligible, therefore we assume that all 3 stars are located at the same place,
and the total mass, luminosity and spectrum of the system is made up of the mass, luminosity and spectrum of the individual components.
 Parameters of the system $\sigma$~Ori~AB are summarized in Table~\ref{tab:3stars}. The reasons for choosing this system are described in Section~\ref{sec:param}.
Two dust materials are considered: the astronomical silicates and graphite (Draine, 2003).
$\langle Q_{rp}\rangle$ and corresponding $\beta(a)$ can be calculated for any radius $a$.
Let us note that here we do not consider a porosity of dust grains, although porosity may influence the $\beta(a)$-curve as shown by Kimura \& Mann~(1999) and investigated in detail by Kirchschlager \& Wolf (2013).

\begin{figure*}
\includegraphics[scale=0.7]{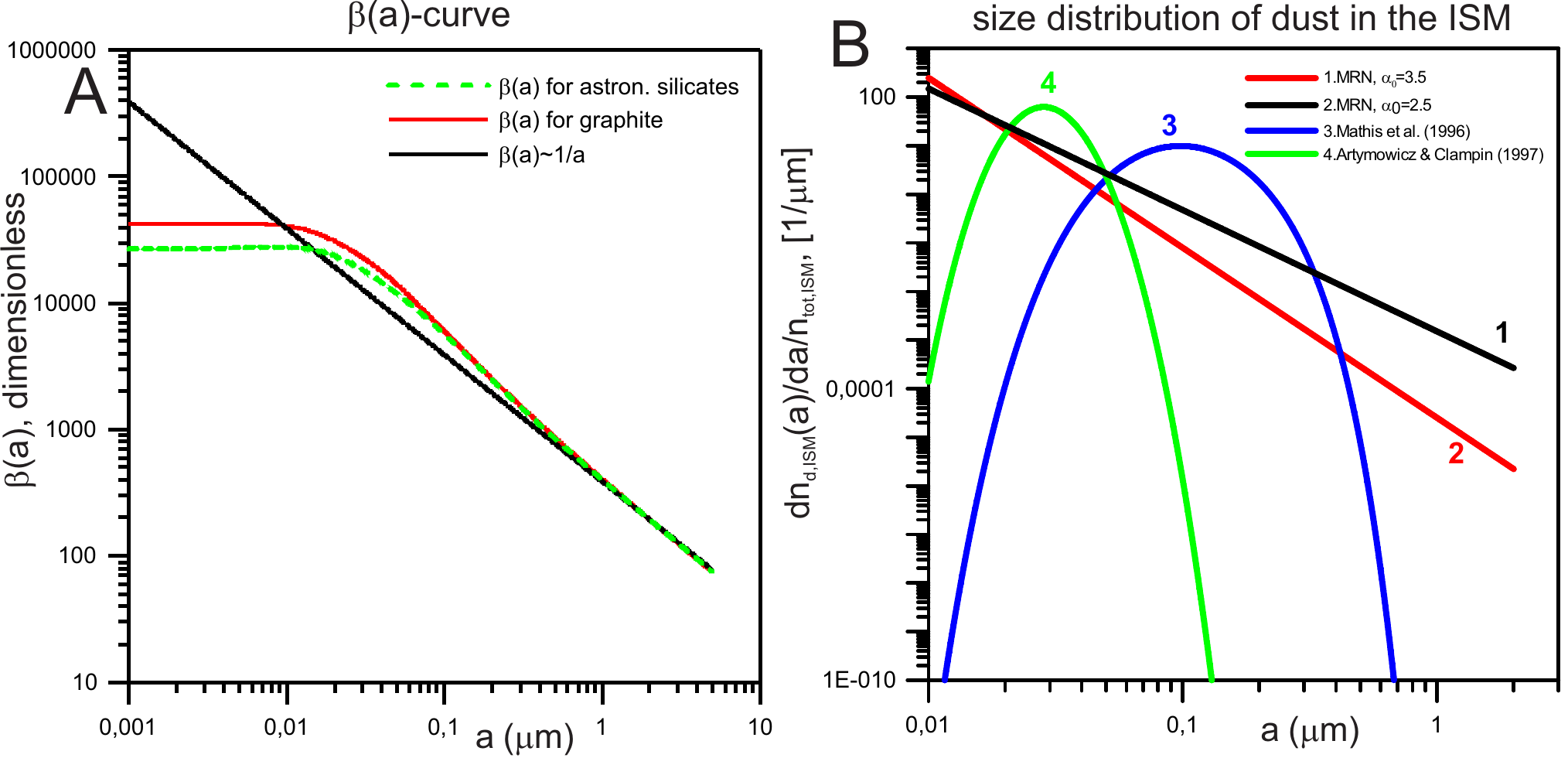}
\centering
\caption{A. $\beta(a)$ calculated numerically for graphite (solid red curve) and silicates (dashed green curve) by using parameters of the system $\sigma$~Ori~AB (see text for details). Black curve corresponds to $\beta(a)\sim1/a$ obtained under
geometrical optics approach.
B. Four normalized dust size distribution in the interstellar medium (see section~\ref{sec:FISM} for details).
}
\label{fig:beta_nism}
\end{figure*}

In literature it is commonly assumed that $\langle Q_{rp}\rangle \approx 1$ that is correct for the geometrical optics approach that, in turn,
is valid only for large dust grains. In this case $\beta(a)\sim 1/a$.
Fig.~\ref{fig:beta_nism}~A shows the numerically calculated $\beta(a)$-curve for silicates and graphite.
It is seen that the geometrical optics approximation works well for $a\geq 0.7~\mu m$.

\subsection{The interstellar dust size distribution}
\label{sec:FISM}

To analyze the dust distribution in the astrospheres one needs to know the size distribution of the interstellar dust in the undisturbed
interstellar medium. Here we consider several typical examples.
 Mathis, Rumpl \& Nordsieck (1977) have suggested the power-law size distribution of dust grains (now it is called the ``MRN'' distribution).
 In this case the ISM number density of dust grains with radius $a\in[a-da/2; a+da/2]$ is
\[
 dn_{d,ISM}(a)=C_{ISM} \cdot a^{-\alpha_0}da,
\]
where $\alpha_0 \in [2.5; 3.5]$.
 Normalization coefficient $C_{ISM}$ can be found from the common assumption that the gas to dust mass ratio in the ISM is 100 (see, e.g., Frisch et al., 1999).
 Note that $[C_{ISM}]=L^{\alpha_0-4}$, where $L$ is measured in units of length.
 There are several works with updated size distribution of dust in the ISM, e.g., Mathis (1996) have proposed
 an advanced expression such as:
 \[
  dn_{d,ISM}(a) =C_{ISM} \cdot a^{-\alpha_0} \exp (-(\alpha_1 a + \frac{\alpha_2}{a}+ \alpha_3 a^2)) da.
 \]
This form resembles a power-law with an exponential cutoff at large sizes (through $\alpha_1$ and $\alpha_3$) and
small sizes (through $\alpha_2$).
Mathis (1996) proposed the following set of coefficients: $\alpha_0=3.5$, $\alpha_1=0.00333$~$(\mu m)^{-1}$,
$\alpha_2=0.437$~$\mu m$, $\alpha_3=50$~$(\mu m)^{-2}$ (hereafter we will call this distribution as ``Mat.96'').
Artymowicz \& Clampin (1997) used the same formula with the different coefficients: $\alpha_1=300$~$(\mu m)^{-1}$ and
$\alpha_2=0.343$~$\mu m$ (hereafter we will call this distribution as ``Artym.97'').

Let us introduce a function $F_{ISM}(a)$ such as:
\[
d n_{ISM}(a)=C_{ISM}\cdot F_{ISM}(a)da,
\]
$F_{ISM}$ is a function from MRN or ``Mat.96'' or ``Artym.97'' size distributions of dust in the ISM.
Therefore, the total dust number density in the ISM is the following:
\begin{equation}
  n_{tot,ISM}=\int_{a_{min}}^{a_{max}} C_{ISM}\cdot F_{ISM}(a)da,
\end{equation}
where $a_{min}$ and $a_{max}$ are the considered minimum and maximum grains's radius in the ISM.
Fig.\ref{fig:beta_nism}~B shows the four considered dust size distributions $dn_{d,ISM}(a)/da$ in the ISM normalized to the total
number density $n_{tot,ISM}$.

\subsection{Dimensionless formulation and parameters}
\label{sec:param}

 Dimensionless formulation of the problem is useful and convenient because it allows to analyze dependence of solution on the dimensionless parameters.
  Dimensionless parameters are constituted from the dimensional ones and, in accordance with $\pi$-theorem (Buckingham, 1914; Sedov, 1993), their number is less than the number of
  dimensional parameters.
  Let us choose the following characteristic dimensional parameters for scaling:
  \begin{itemize}
    \item the character distance is usually taken as a stand-off distance from the star to the astropause at the nose direction that is
    $L^*=\sqrt{\frac{\dot{M} v_\infty}{4\pi \rho_{ISM}V_{ISM} ^2}}$, where $\dot{M}$ is the stellar mass loss rate, $v_\infty$ is the
    stellar wind velocity, $\rho_{ISM}=1.4 m_p n_{p,ISM}$ (H mass fraction is assumed to be $X=0.715$, Asplund et al. 2009, hence,
    the mean mass per ion is $\mu_H=1 / X=1.4$ and $\rho_{ISM}=\mu_H m_p n_{p,ISM}$);
    \item the character velocity is $V_{ISM}$;
    \item the character number density is $n_{p,ISM}$ and corresponding character density is $\rho_{ISM}$;
    \item the character temperature is $T_{ISM}$;
    \item the character magnetic field is $B_{ISM}$.
  \end{itemize}
All dimensional variables can be scaled by these quantities. Note that time is scaled as follows:
\[
  \hat{t}=t \cdot \frac{V_{ISM}}{L^*}.
\]
Therefore equation~(\ref{eq:motion_beta}) can be represented in the following dimensionless form (variables with a ``hat'' are dimensionless):
\begin{eqnarray}\label{eq:motion_br}
 \frac{d\hat{\textbf{v}}_d}{d\hat{t}} =\sgn(\beta(a)-1) \, \frac{L_1}{L^*}\cdot \frac{\textbf{e}_r}{\hat{r}^2} +
\frac{L^*}{L_2} \cdot (\hat{n}_p \hat{T}_p \hat{G}_{pen}(s_p,s_e,s_n)) \cdot \textbf{e}_{rel} + \\
 + \frac{L^*}{L_3} \cdot ((\hat{\textbf{v}}_d-\hat{\textbf{v}}_p)\times \hat{\textbf{B}}), \nonumber
\end{eqnarray}
where the following character length of each force are introduced:
\begin{align}\label{eq:L123}
 L_1&=\frac{G M_s}{ V_{ISM}^2}\cdot |\beta(a)-1|, \nonumber \\
 L_2 &= \frac{2 \sqrt{\pi} V_{ISM}^2 \rho_d}{3 n_{p,ISM} k_b T_{ISM}}\cdot a, \\
 L_3&=\frac{4\pi \rho_d c_0 V_{ISM}}{3 U_d B_{ISM}} \cdot a^2  \nonumber
\end{align}
These lengths can be calculated from known dimensional parameters of the dust grains and the ISM.
Note that variables $s_p$, $s_e$ and $s_n$ can be represented as ($i=\{p,e,n\}$):
\[
 s_{i}=\sqrt{\frac{m_{i}\hat{v}_{rel}^2}{2k_b \hat{T}_p}}\cdot\frac{V_{ISM}}{\sqrt{T_{ISM}}},
\]
and
\begin{equation}\label{eq:Mism}
   \frac{V_{ISM}}{\sqrt{T_{ISM}}} =\sqrt{ \frac{2k_b \gamma}{1.4 m_p}} \cdot M_{ISM},
\end{equation}
where, $\gamma=5/3$ is the ratio of specific heat fluxes for monoatomic gas, and $M_{ISM}$ is the gasdynamic Mach number of the ISM plasma.
The boundary condition for the dimensionless dust velocity in the interstellar medium is
\[
 \hat{\textbf{v}}_d = -1\cdot \textbf{e}_z,
\]
where $\textbf{e}_z$ is a unity vector opposite to $\textbf{V}_{ISM}$.

There are 3 independent dimensionless parameters, which determine the dust motion in the astrosphere outside of the astropause: $L_1/L^*$, $L^*/L_2$ and $L^*/L_3$
(it is assumed that the Mach number $M_{ISM}$ is fixed and known from the plasma parameters). All other terms at the right part of
the motion equation~(\ref{eq:motion_br}) such as $\hat{r}$, $\hat{n}_p$, $\hat{T}_p$, $\hat{G}_{p,e,n}$, $\hat{v}_{rel}$ and $\hat{B}$ are
of the order of unity due to chosen way of normalization to the interstellar parameters. Note that this is not correct within the region
occupied by the stellar wind, where plasma parameters may be significantly different from the interstellar ones.

To calculate ($L^*, L_1, L_2, L_3$) one needs to know (or estimate somehow) the following dimensional parameters:
\begin{itemize}
  \item ISM parameters: $n_{p,ISM}$, $V_{ISM}$, $T_{ISM}$, $B_{ISM}$;
  \item stellar parameters: $L_s$, $M_s$, $v_{\infty}$, $\dot{M}$;
  \item dust parameters: $U_d$, $\rho_d$, radius $a$.
\end{itemize}
Then it is possible to calculate dimensionless parameters $L_1/L^*$, $L^*/L_2$ and $L^*/L_3$ for all typical sizes of dust grains.
If one of dimensionless parameters is much larger than other for all dust grains in the considered region (e.g. where the arc in observed) then it is possible to keep only one term in equation~(\ref{eq:motion_br}).
Let us consider the three limiting cases, when each of the forces dominates over other.
\begin{description}
  \item[1. The central force (a combination of $\textbf{F}_{rad}$ and $\textbf{F}_g$) dominates:]
  \[
   \begin{cases}
    \frac{L^*}{L_2} \ll \frac{L_1}{L^*} \\
    \frac{L^*}{L_3} \ll \frac{L_1}{L^*} \\
   \end{cases} \Rightarrow
    \begin{cases}
     L^* \ll \sqrt{L_1 L_2} \\
     L^* \ll \sqrt{L_1 L_3} \\
   \end{cases}
  \]
  \item[2. The drag force dominates:]
 \[
   \begin{cases}
    \frac{L_1}{L^*} \ll \frac{L^*}{L_2} \\
    \frac{L^*}{L_3} \ll \frac{L^*}{L_2} \\
   \end{cases} \Rightarrow
    \begin{cases}
     L^* \gg \sqrt{L_1 L_2} \\
     L_3 \gg L_2 \\
   \end{cases}
  \]

  \item[3. The electromagnetic force dominates:]
\[
   \begin{cases}
    \frac{L_1}{L^*} \ll \frac{L^*}{L_3} \\
    \frac{L^*}{L_2} \ll \frac{L^*}{L_3} \\
   \end{cases} \Rightarrow
    \begin{cases}
      L^* \gg \sqrt{L_1 L_3} \\
      L_2 \gg L_3 \\
   \end{cases}
  \]
\end{description}

 As mentioned above a comparison of $L_1/L^*$, $L^*/L_2$ and $L^*/L_3$ makes sense only outside of the astropause. Therefore, let us consider the additional criteria, which guaranties that dust grains do not penetrate inside the astropause.
As it will be shown in the next Section, if the radiation pressure force dominates then there is a dust cavity (where dust can not penetrate) located at distance $2\cdot L_1(a)$
  from the star in the nose direction. By the definition of $L^*$, distance from the star to the astropause in the nose direction is $L^*$. Therefore, additional condition $2\cdot L_1(a)>L^*$
  guarantees that dust grains with radius $a$ do not penetrate inside the astropause.

For the second case, when the drag force dominates, we consider the simplified form of the motion equation in order to get the additional criteria ensuring that the dust grain does not penetrate inside the astropause.
Namely, let us consider the drag force caused only by an interaction with protons. As the first approximation we can assume $\hat{G}_p\sim s_{p}\sim v_{rel}$. Therefore the motion equation can be rewritten as:
\begin{equation}\label{eq:case2}
 \frac{d\hat{\textbf{v}}_d}{d\hat{t}} = \frac{L^*}{L_2} \left( \hat{n}_p \hat{T}_p \frac{8}{3 \sqrt{\pi}} \sqrt{\frac{m_p}{2k_b T_p}} \right) \cdot \textbf{v}_{rel}.
\end{equation}
Let us introduce a function $\hat{H}(\hat{n}_p, \hat{T}_p)$:
\begin{eqnarray}
  \hat{H}(\hat{n}_p, \hat{T}_p) = \hat{n}_p \hat{T}_p \frac{8}{3 \sqrt{\pi}} \sqrt{\frac{m_p}{2k_b T_p}} V_{ISM} = \nonumber \\
  = \hat{n}_p \sqrt{\hat{T}_p} \frac{8}{3 \sqrt{\pi}} \sqrt{\frac{\gamma}{1.4}} M_{ISM} \approx \hat{n}_p \sqrt{\hat{T}_p} \cdot 1.64 \cdot M_{ISM}, \nonumber
\end{eqnarray}
here we substitute $T_p=\hat{T}_p \cdot T_{ISM}$ and use equation~(\ref{eq:Mism}).
Let us consider the motion equation~(\ref{eq:case2}) locally under assumption of constant plasma parameters. In this simplified case a projection of the equation~(\ref{eq:case2})
 to $\textbf{e}_{rel}=(\textbf{v}_p-\textbf{v}_d)/|\textbf{v}_p-\textbf{v}_d|$ can be represented as:
\begin{equation}
 \frac{d(\hat{v}_d-\hat{v}_p)}{d\hat{t}} = -\frac{L^*}{L_2} \cdot \hat{H} \cdot (\hat{v}_d-\hat{v}_p).
\end{equation}
This equation can be solved analytically:
\[
   \hat{v}_d-\hat{v}_p = \hat{v}_{rel,0} \cdot \exp (-\frac{L^*}{L_2} \hat{H} \cdot \hat{t}),
\]
where $\hat{v}_{rel,0}$ is an integration constant. It is seen from the solution that $\hat{v}_d \approx \hat{v}_p$ for $\hat{t}$ such as $\hat{t} > L_2/(L^* H)$.
Outside of the astropause $H\sim 1-10$ (it depends on $M_{ISM}$). Therefore, if $L_2 \ll L^*$ then $\hat{v}_d$ becomes close to $\hat{v}_p$ at very small dimensionless time, so we can assume that it happens instantly. Therefore, under condition of $L_2 \ll L^*$ the dust will be ``frozen'' into the gas motion and it can not penetrate inside the astropause.

For the third case, the electromagnetic force dominates. And the motion of a dust particle can be represented as
the gyrorotation around the magnetic field line frozen into the plasma and the motion of the guiding center.
The guiding center is moving due to $\textbf{E} \times \textbf{B}$ drift (velocity of this drift is equal to the plasma velocity) and
possible gradient drift caused by variations of the magnetic and electric fields (electric field is $\textbf{E}=-\textbf{v}_p \times \textbf{B}$).
If gyrorotation is very fast (i.e. dimensionless gyroradius is much smaller than unity) and there are no extreme gradients of plasma parameters, then the dust grains move together with plasma and do not penetrate through the astropause.
Dimensionless gyroradius or Larmor radius is the following:
\[
 \hat{r}_{gyr}=\frac{m_d v_{d,\perp} c_0}{q B}\cdot \frac{1}{L^*}=\frac{4\pi\rho_d c_0 a^2 v_{d,\perp}}{3 U_d B}\cdot \frac{1}{L^*} = \frac{L_3}{L^*} \frac{\hat{v}_{d,\perp}}{\hat{B}},
\]
 where $v_{d,\perp}$ is the dust velocity component perpendicular to the magnetic field vector. Outside of the astropause $v_{d,\perp} \sim 0-1$ and $\hat{B} \sim 1$.
 Thus, if $L_3 \ll L^*$ then $\hat{r}_{gyr} \ll 1$ and this guarantees that dust grain is moving together with plasma component and flow around the astropause.


Therefore, in additional to three criteria formulated above one needs to compare $L^*$ with $L_i$ ($i=1,2,3$) and check whether dust penetrates inside
the astropause or not.

\begin{figure*}
\includegraphics[scale=0.8]{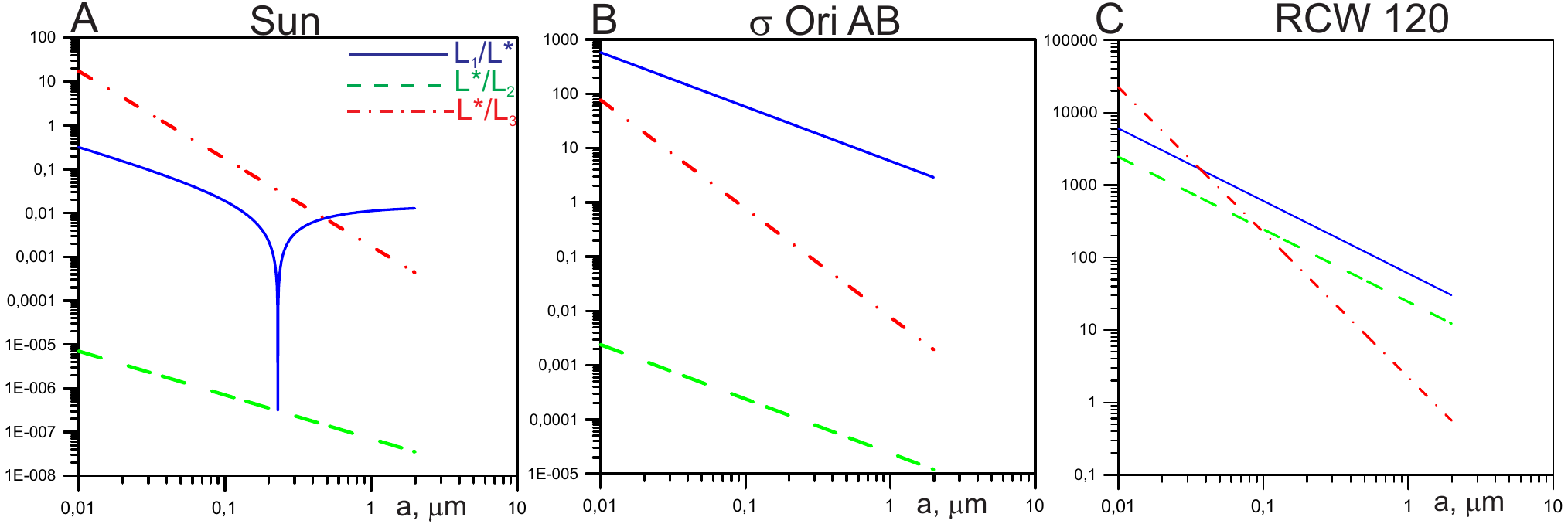}
\centering
\caption{Dimensionless parameters ($L_1/L^*$, $L^*/L_2$ and $L^*/L_3$) calculated for three stars: Sun (A), $\sigma$~Ori~AB (B) and
CD-38$^{\circ}$11636 in RCW 120 (C). In the case of the Sun $\beta(a)>1$ for small grains with $a<0.233$~$\mu m$ and $\beta(a)<1$
for large grains with $a>0.233$~$\mu m$, therefore a function $L_1(a)$ that is proportional to $|\beta(a)-1|$ has an inflection point at $a=0.233$~$\mu m$.
}
\label{fig:3starsL}
\end{figure*}

If all $L_i$ ($i=1,2,3$) and $L^*$ are known then it is possible to test these criteria and determine, which forces are important.
For example, we calculated ($L^*, L_1, L_2, L_3$) for three certain stars: Sun, $\sigma$~Ori~AB (Ochsendorf et al., 2014a; Sim\'{o}n-D\'{i}az et al., 2015) and CD-38$^{\circ}$11636 in RCW 120
(Mackey et al., 2015, 2016, Ochsendorf et al., 2014b).
 Corresponding dimensional parameters are summarized in Table~\ref{tab:3stars}.
Plot~\ref{fig:3starsL} shows $L_1/L^*$, $L^*/L_2$ and $L^*/L_3$ calculated as functions of dust grains radius in the geometrical optics approach i.e. with $\langle Q_{rp}\rangle \approx 1$.
It is seen that for the Sun
$L^*/L_3$ is several order of magnitude larger than $L^*/L_2$ for all grains radii and at least 10 times larger than $L_1/L^*$ for dust grains with $a\leq 0.25$~$\mu m$.
For larger dust grains $L_1/L^*$ becomes to be comparable or even larger than $L^*/L_3$.
 It is seen that $L^*>L_3$ only for small grains with $a\leq 0.05$~$\mu m$. These grains do not penetrate through the
 astropause and their motion in the interstellar plasma is determined only by the electromagnetic force.

  For system $\sigma$~Ori~AB the situation is different and the radiation pressure dominates
for dust grains with all sizes due to high luminosity of the star, weak stellar and strong interstellar winds. $L_1>L^*$ for all considered dust grains radii, therefore these grains
 do not cross the astropause. The large dust grains for which $2\cdot L_1<L^*$, in principle, could penetrate inside the astropause, and the drag force may influence their dynamics.

 In the case of H~II region RCW~120 with chosen magnitude of $B_{ISM}$ for dust grains with $a\leq1\, \mu m$ all
three forces are comparable, for large dust grains with $a\geq1\, \mu m$ the electromagnetic force can be neglected, but the radiation
pressure force and the drag force are the same order of magnitude and should be considered together.

\begin{table}
  \caption{Dimensional parameters of the astrospheres for three objects.}
  \label{tab:3stars}
  \renewcommand{\footnoterule}{}
 \begin{center}
 \begin{minipage}{\textwidth}
  \begin{tabular}{c|ccc}
\hline
   & Sun$^{(a)}$ & $\sigma$~Ori~AB$^{(b)}$ &  RCW 120$^{(c)}$  \\
\hline
  $log(L_s/\lsun)$ & 1 & 4.88 & 5.07  \\
  $M_s$ ($\msun$) & 1 & 45 & 30  \\
  $T_s$ (kK) & 5.777 & 35; 31; 29 & 37.5  \\
  $\dot{M}$ ($\msun$ year$^{-1}$) & $ 10^{-14}$ & $2\cdot 10^{-10}$ & $1.55\cdot 10^{-7}$  \\
  $v_{\infty}$ (km s$^{-1}$) & 400 & 1500 & 2000  \\
  $V_{ISM}$ (km s$^{-1}$) & 26.4 & 50 & 4  \\
  $n_{p,ISM}$ (cm$^{-3}$) & 0.1 & 10 & 3000  \\
  $T_{ISM}$ (K) & 8000 & 8000 & 7500  \\
  $B_{ISM}$ ($\mu$G) & 4.4 & 3$^{(d)}$ & 3$^{(d)}$ \\
  $\rho_d$ (g cm$^{-3}$) & 2.5 & 2.5 & 2.5  \\
  $U_d$ (V) & 0.75$^{(e)}$ & 0.75 & 0.75  \\
\hline
\end{tabular}
\end{minipage}
\end{center}
  $\msun$ is the mass of the Sun;
  $\lsun$ is the luminosity of the Sun.
  $^{(a)}$ Izmodenov \& Alexashov (2015);
  $^{(b)}$ Ochsendorf et al., 2014a; Sim\'{o}n-D\'{i}az et al. (2015), the system $\sigma$~Ori~AB consists of three stars, their individual parameters are described by Sim\'{o}n-D\'{i}az et al. (2015).
  The total mass and luminosity of $\sigma$~Ori~AB are found as a sum of mass and luminosity of three components, and effective temperature of three components are provided;
  $^{(c)}$ Mackey et al., 2015, 2016;
  $^{(d)}$ typical magnetic field strength in the warm phase of the ISM (Troland \& Heiles 1986;
Harvey-Smith et al. 2011)
  $^{(e)}$ Gr$\ddot{u}$n \& Svestka (1996). For these estimations we keep the same value of $U_d$ for all stars, although in reality it may depends on
  the stellar and isotropic interstellar radiation in the vicinity of the certain star.
\end{table}

In this work we consider only the first limiting case, when the dust motion is determined by the radiation repulsive force.
This case could takes place for the star with weak stellar wind and high luminosity, when the dust is swept out by the stellar radiation pressure
much further than the position of the bow shock. Therefore the arc is formed only due to the stellar radiation and has nothing to do with the bow shock.
In our calculations where it is necessary we use dimension parameters of system $\sigma$~Ori~AB, because the radiation pressure force dominates for this system.

\section{Analytical solution for the dust density distribution}
\label{sec:cold_model}

\subsection{Trajectories and dimensionless number density distribution for certain grain radius $a$}
 Let us consider the case when the drag and electromagnetic forces are neglected compared to the central force ($\textbf{F}_g+\textbf{F}_{rad}$).
 Therefore the dimensionless equation of motion is the following:
 \begin{equation}
   \frac{d\hat{\textbf{v}}_d}{d\hat{t}} =\sgn(\beta(a)-1) \,\frac{L_1}{L^*}\cdot \frac{\textbf{e}_r}{\hat{r}^2}.
 \end{equation}
 As it is seen from Fig.~\ref{fig:beta_nism}~A, in the considered case $\beta(a)\gg1$ and, hence:
  \[
    L_1(a) = \frac{3L_s \langle Q_{rp}\rangle }{16 \pi c_0 \rho_d V_{ISM}^2} \cdot \frac{1}{a}.
  \]
  In this case the total force is the repulsive force and the arc of dust around the star can be formed.
 Note also that the problem is axisymmetric.

Previously, when we considered the general situation with three forces (eq.~(1)) it was convenient to choose the character distance $L^*$ as the stand-off distance to the astropause (as we did in section 2.3).
 Now the situation is much more simpler, we have only one force and solution does not depend on the astropause anymore. Therefore in this limiting case it is convenient
 to choose new character distance, which allow us to rewrite our motion equation in appropriate form.
Namely, for each certain dust grain radius let us to choose the character distance $D^*(a)=L_1(a)$ (instead of $L^*$). In this case the equation becomes even simpler:
\begin{equation}\label{eq:cold_model}
   \frac{d\hat{\textbf{v}}_d}{d\hat{t}} = \frac{\textbf{e}_r}{\hat{r}^2}.
 \end{equation}
This dimensionless equation does not depend on grain's radius $a$, hence, its dimensionless solution is the same for all dust grains.

\begin{figure*}
\includegraphics[scale=0.7]{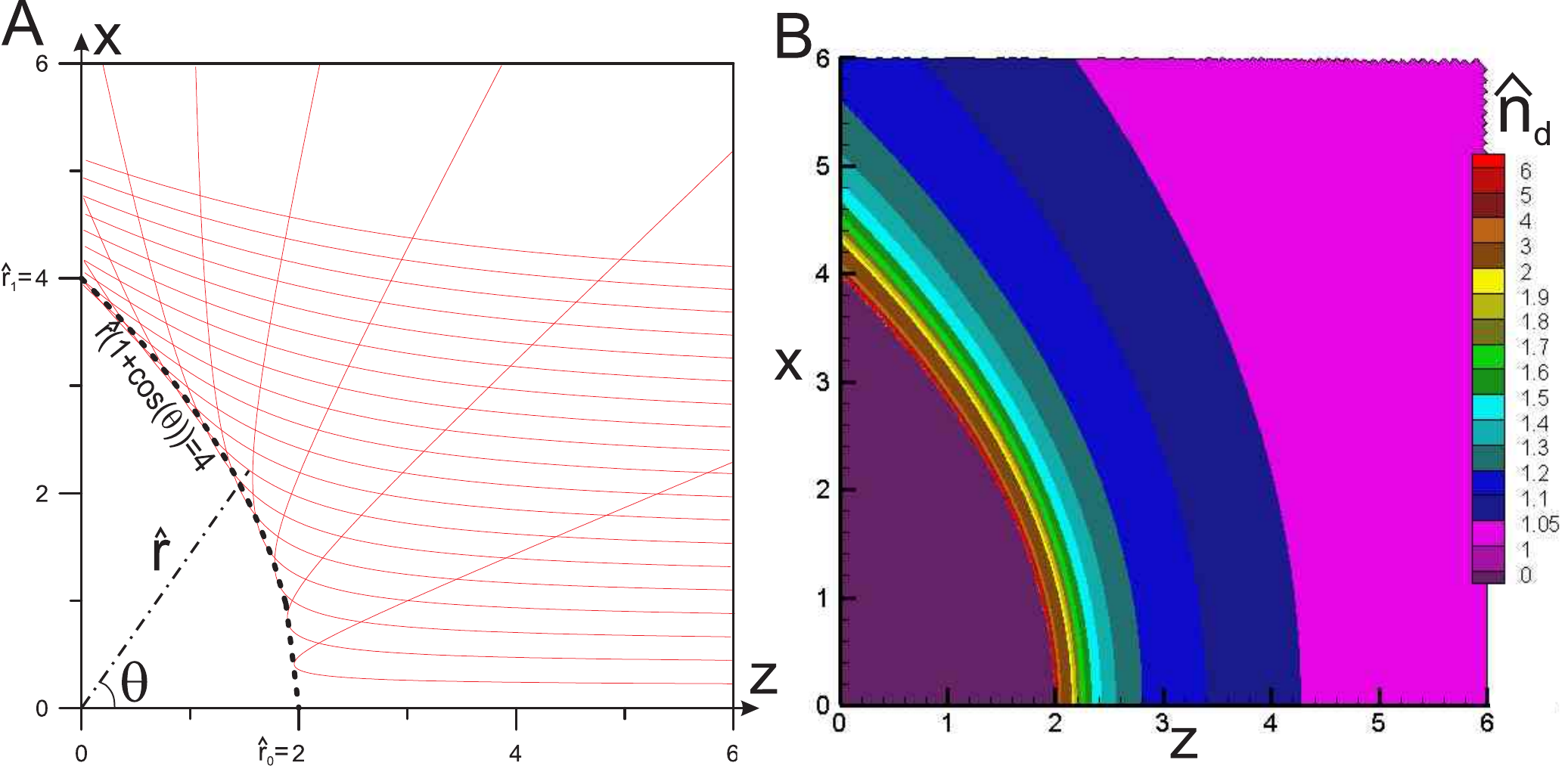}
\centering
\caption{A. Hyperbolic trajectories that are solution of dust motion equation~\ref{eq:cold_model} are shown.
B. Distribution of dimensionless number density of dust corresponding to classical cold model (see equation~\ref{eq:nd_cold_model}).
Note that the dust number density at parabolic boundary of the dust cavity ($\hat{r}(1+cos(\theta)=4$) is infinite as it is follows from eq.~(\ref{eq:nd_cold_model}), but
infinite values can not be visualized by color scheme, so maximum value of $\hat{n}_d=6$ is marked.
}
\label{fig:n_cold}
\end{figure*}

Solution of the equation~(\ref{eq:cold_model}) is the hyperbolic trajectories (see, e.g., Landau \& Lifshitz, 1969).
If all dust grains have the same velocity at infinity then only two hyperbolic trajectories pass through each point ($\hat{r}, \theta$) with
$\hat{r}(1+cos(\theta))>4$ (here $\theta$ is an angle counted from $z$-axis, see Fig.~\ref{fig:n_cold}~A). Parabola defined by equation $\hat{r}(1+cos(\theta))=4$ is a boundary of, so called, forbidden cavity, where dust particles can not
penetrate.
Fig.~\ref{fig:n_cold}~A shows the trajectories of dust grains obtained as the solution of the dust motion equation~(\ref{eq:cold_model}).

Consideration of dust continuity equation in the case of the cold model allows to obtain the following analytical solution for the dimensionless number density of dust (see, e.g., Axford 1972, Lallement et al., 1985):
\begin{equation}\label{eq:nd_cold_model}
  \hat{n}_d(\hat{r},\theta) = \frac{(1+\sqrt{A})^2}{4\sqrt{A}}+\frac{(1-\sqrt{A})^2}{4\sqrt{A}}=\frac{1+A}{2\sqrt{A}},
\end{equation}
where $A=1-4/[\hat{r}(1+cos(\theta))]$.
Fig.~\ref{fig:n_cold}~B presents the dimensionless dust distribution described by equation~(\ref{eq:nd_cold_model}).
Note that the number density at the curve $A=0$ (that is the forbidden cavity's boundary) is infinite as it is seen from equation~(\ref{eq:nd_cold_model}).
This arc has a parabolic shape, and the dimensionless distance from the star to the arc in the nose direction ($\theta=0^{\circ}$) is
$\hat{r}_0=2$ and in the side direction ($\theta=90^{\circ}$) is $\hat{r}_1=4$.
The problem is axisymmetric, i.e. a three-dimension (3D) number density distribution can be obtained by rotation of the presented one around $z$-axis.
Note that the solution of equation~(\ref{eq:nd_cold_model}) exists for all magnitudes of angle $\theta$ from 0 to 180$^{\circ}$. Here we present the results only for the upwind hemisphere with $z>0$
($\theta \leq 90^{\circ}$) just because of most of the infrared arcs around the stars are observed in the nose part of the astrosphere.

In order to obtain the dimensional solution for chosen grain's with radius $a\in[a-da/2; a+da/2]$ from the presented dimensionless solution we need just to rescale the distances $r=\hat{r}\cdot D^*(a)$ and multiply the number density by $dn_{d,ISM}(a)$ that is the ISM number density of dust grains with chosen
 interval of radius, i.e.
\[
 dn_d(r,\theta,a)=\hat{n}_d\left(\frac{r}{D^*(a)},\theta\right)\cdot dn_{d,ISM}(a).
\]
In the next section we consider the dust number density distribution for the mixture of dust with different radii.

\subsection{Dust number density distribution for mixture of grains with different sizes}


To obtain the dust number density in the astrosphere for the mixture of dust grains with different radii we need to integrate the dimensional solutions
over radius $a$, namely:
\begin{equation}\label{eq:n_mix}
  n_{d,mix}(r,\theta) = \int_{a_{min}}^{a_{max}} \hat{n}_d\left(\frac{r}{D^*(a)}, \theta \right) \, C_{ISM}\cdot F_{ISM}(a)da,
\end{equation}
where $a_{min}$ and $a_{max}$ are the minimum and maximum dust grains radii considered in the ISM.

The dust distribution for each certain grain's radius has a distinct parabolic arc with infinite number density at a distance $r_0=2\cdot D^*(a)$ from the star
in the upwind direction ($\theta=0^{\circ}$) and at distance $2r_0$ in perpendicular direction ($\theta=90^{\circ}$).
The question is whether the total number density distribution with different grain's radius has the similar arc and where the arc is located?

It is convenient to introduce the characteristic scale, which does not depend on the grain's radius.
Let's choose the following distance: $L_0=D^*(a_0)= G M_s (\beta(a_0)-1)/V_{ISM}^2$ with $a_0=1~\mu m$.
For system $\sigma$~Ori~AB $L_0=0.03$~pc.
Hereafter we use this scale for presentation of all results obtained for the mixture of dust with different sizes, i.e. all distances in the plots are
scaled as $\hat{r}=r/L_0$.
Note that:
\[
  \frac{r}{D^*(a)} = \hat{r} \cdot \frac{L_0}{D^*(a)} = \hat{r} \cdot \frac{\beta(a_0)-1}{\beta(a)-1}.
\]

In the simplest case with $\beta(a)\sim 1/a$ (geometrical optics) and the classical MRN size distribution of dust in the ISM the integral~(\ref{eq:n_mix}) can be found analytically and it is finite for all points $(r,\theta)$ (see Appendix~\ref{app_A} for details). For the cases with
numerical values of $\beta(a)$ and/or with more complicated size distribution in the ISM the integral can be calculated numerically by, e.g.,
Gauss integration method.


\begin{figure*}
\includegraphics[scale=0.9]{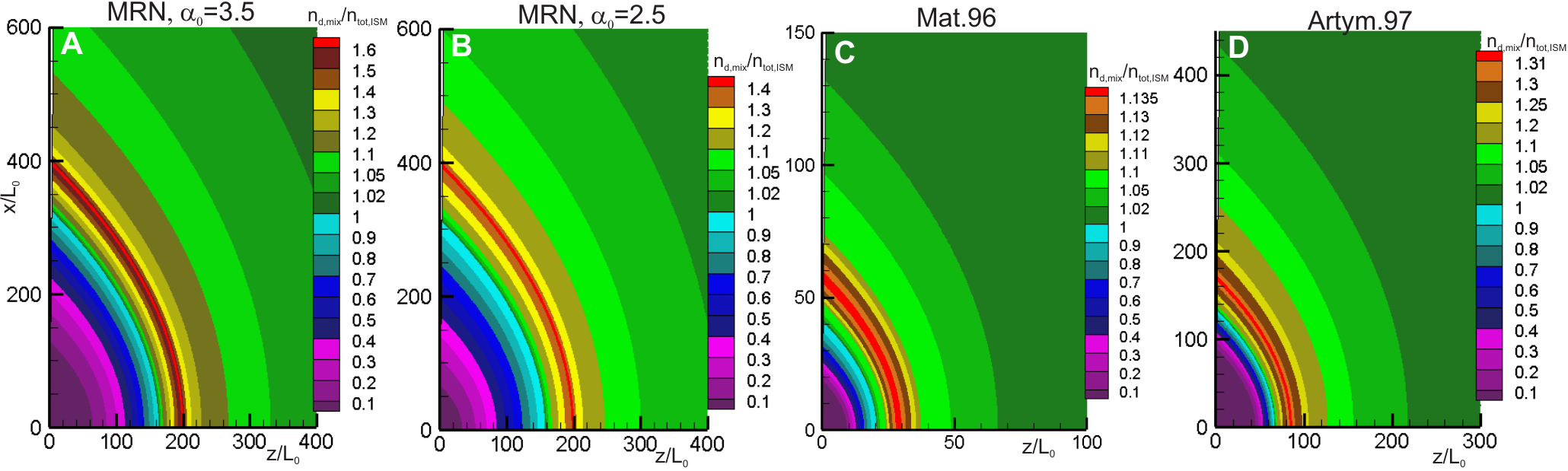}
\centering
\caption{2D distributions of normalized dust number density obtained for the mixture of dust grains with different dust size distributions in the ISM.
These results are obtained under the geometrical optics approach with $\beta(a)\sim 1/a$.
}
\label{fig:n_mix}
\end{figure*}

Fig.~\ref{fig:n_mix} demonstrates the normalized dust number density $n_{d,mix}/n_{tot,ISM}$ around the star obtained for
four considered size distributions in the ISM with $a_{min}=10^{-2}~\mu m$ and $a_{max}=2~\mu m$ and with $\beta(a)\sim 1/a$.
 It is seen that for the MRN distributions (Plots A and B) the arc with enhanced dust number density exists at distance $z/L_0=200=2/a_{min}$ ($[a_{min}]=\mu m$). This is because of a power-law distribution in the ISM where the smallest grains have the largest
 number density. However, in contrast to the number density distribution for individual grain's radius, the total number density
 for mixture of dust grains at the arc is finite. This is because of a singularity at the arc ($A=0$) in equation~(\ref{eq:nd_cold_model})
 is integrable over $a$ (see Appendix~A).
 Note that the shape of the arc is the same for all considered dust size distributions in the ISM.
 There are some differences in the distributions obtained with $\alpha_0$=3.5 and 2.5. Namely, in the case of $\alpha_0=3.5$ the outer arc
 is brighter and a cavity with small amount of dust near the star is wider as compared with the case of $\alpha_0=2.5$.
 This is explained by the following. In the ISM with $\alpha_0=3.5$ there are more small grains and less large grains compared with the case of $\alpha_0=2.5$ (see Fig.~\ref{fig:beta_nism}~B). Therefore in the astrosphere with $\alpha_0=3.5$ we have a smaller amount of dust close to the star (where only big grains can penetrate)
 and larger amount of dust far away from the star (where the small dust grains dominate).
 For two other size-distributions (Plots C and D) the arcs with maximum number density are located closer to the star compared with the MRN cases,
 because for both ``Mat.96'' and ``Artym.97'' distributions maximum of number density corresponds to certain grain's radius $a_c>a_{min}$ (see Fig.~\ref{fig:beta_nism}~B), therefore
 these grains penetrate closer to the star. Dust number density in the arc is smaller in the case of ``Mat.96'' (Plot C) than in the case
 of ``Artym.97'' (Plot D) in accordance with height of maximum number density in the ISM (see Fig.~\ref{fig:beta_nism}~B).

\begin{figure*}
\includegraphics[scale=0.9]{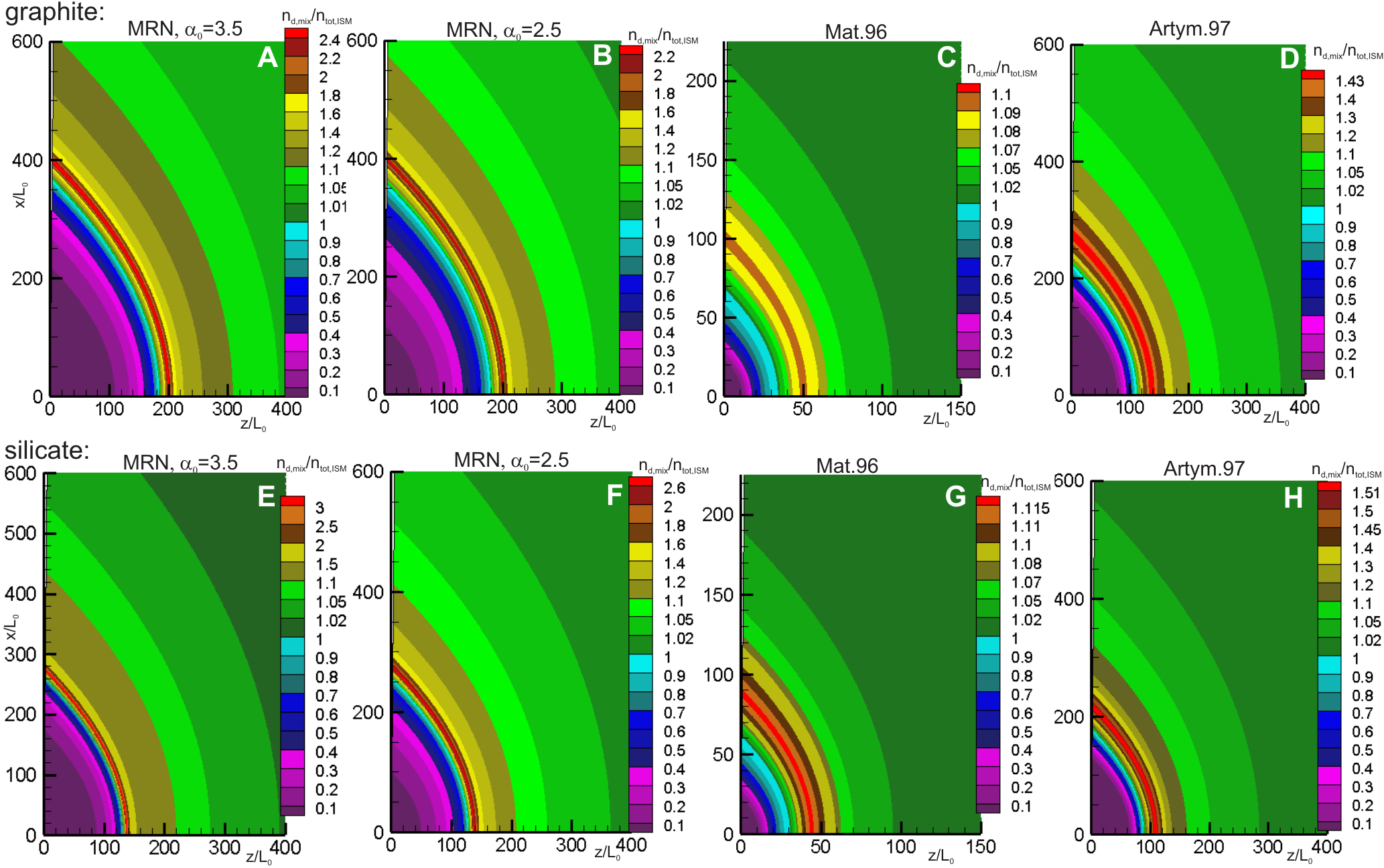}
\centering
\caption{2D distributions of normalized dust number density obtained for the mixture of dust grains with four different dust size distributions in the ISM.
These results are obtained with numerically calculated $\beta(a)$-values for graphite (plots~A-D) and silicates (plots E-H) separately.
}
\label{fig:n_mix_bet}
\end{figure*}


 Fig.~\ref{fig:n_mix_bet} shows the dust number density distributions obtained with numerically calculated $\beta$-curve (see Fig.~\ref{fig:beta_nism}~A)
 for graphite and silicates. It is seen that the arc with the same parabolic shape exists for all considered variants. For all considered size distribution
 in the ISM the arc is closer to the star for silicates than for graphite, because $\beta$ is larger for graphite and, therefore, these dust grains are
 swept out further from the star. Magnitude of number density inside the arc strongly depends on the dust size distribution
 assumed in the ISM. The arc is produced by grains with the size $a_{peak}$ that corresponds to maximum number density in the ISM (see Fig.~\ref{fig:beta_nism}~B). E.g. for the MRN size distribution $a_{peak}=0.01$~$\mu m$ (at the considered range of dust sizes from 0.01~$\mu m$
 to 2~$\mu m$),
 for ``Artym.97'' $a_{peak}=0.028$~$\mu m$ and for ``Mat.96'' $a_{peak}=0.1$~$\mu m$.
 Position of the arc is determined by magnitude of $\beta(a_{peak})$ for such grains: for smaller $\beta(a_{peak})$ the arc is located closer to the star.
 With the numerically calculated $\beta$-curve in the case of MRN distribution for graphite the arcs are located at the same place as for
 $\beta(a)\sim 1/a$ (compare Plots A-B in Figures~\ref{fig:n_mix} and \ref{fig:n_mix_bet}) because by coincidence the magnitudes of $\beta$ are the same for $a=0.01$~$\mu m$, which provide the arc in the case of MRN distribution.
 MRN distribution with $\alpha_0$=3.5 provides the largest number density within the arc, while ``Mathis.96'' distribution
 provides the smallest one in accordance with $n_{d,ISM}(a_{peak})$ that is a maximum of the dust number density in the ISM.
 A comparison of Fig.~\ref{fig:n_mix_bet} with Fig.~\ref{fig:n_mix} shows that using a geometrical optics approximation, i.e.
 $\beta(a)\sim 1/a$, leads to underestimation of the dust number density within the arc and a visible shift of the arc.

\section{Calculation of the infrared thermal dust emission intensity maps}

Many astrospheres are observed due to the thermal emission from the interstellar dust grains that is caused by heating of dust grains by the stellar radiation.
Typical wavelength of the observed thermal emission from dust is $\lambda_0=24\,\mu m$.
For the chosen direction the thermal emission intensity is an integral of emissivity ($j$) along the line-of-sight (LOS):
$I(\lambda_0)=\int j(\lambda_0, \textbf{r}(s))ds$, where $s$ is the coordinate along LOS.
The local emissivity is the following:
\begin{eqnarray}\label{emis}
 j(\lambda_0, \textbf{r}) = \int_{a_{min}}^{a_{max}} \pi a^2 \hat{n}(\frac{r}{D^*(a)}, \theta) \cdot C_{ISM} F_{ISM}(a) \cdot \\
  \cdot Q_{abs}(a, \lambda_0) B_{\nu_0}(T_d(a,r)) da,  \nonumber
\end{eqnarray}
where,
$Q_{abs}(a,\lambda_0)$ is the dust absorption coefficient calculated by means of Mie theory, $B_{\nu_0}$ is the Plank function with the dust grain's temperature $T_d(a,r)$:
\[
 B_{\nu_0}(T_d(a,r)) = \frac{2h\nu_0^3/c_0^2}{exp(\frac{h\nu_0}{k_b T_d(a,r)})-1},
\]
where, $h$ is the Plank constant, $c_0$ is the speed of light, $\nu_0=c_0/\lambda_0$.
Here we consider only LOS along $y-$axis, i.e. the astrosphere is observed from the flank (an angle between the LOS and vector $\textbf{V}_{ISM}$
is 90$^{\circ}$). The problem is axisymmetric, therefore,
it is needed to calculate emissivity only in $(zx)$-plane and then the 3D distribution of emissivity is obtained by rotation around $z$-axis.

\begin{figure*}
\includegraphics[scale=0.7]{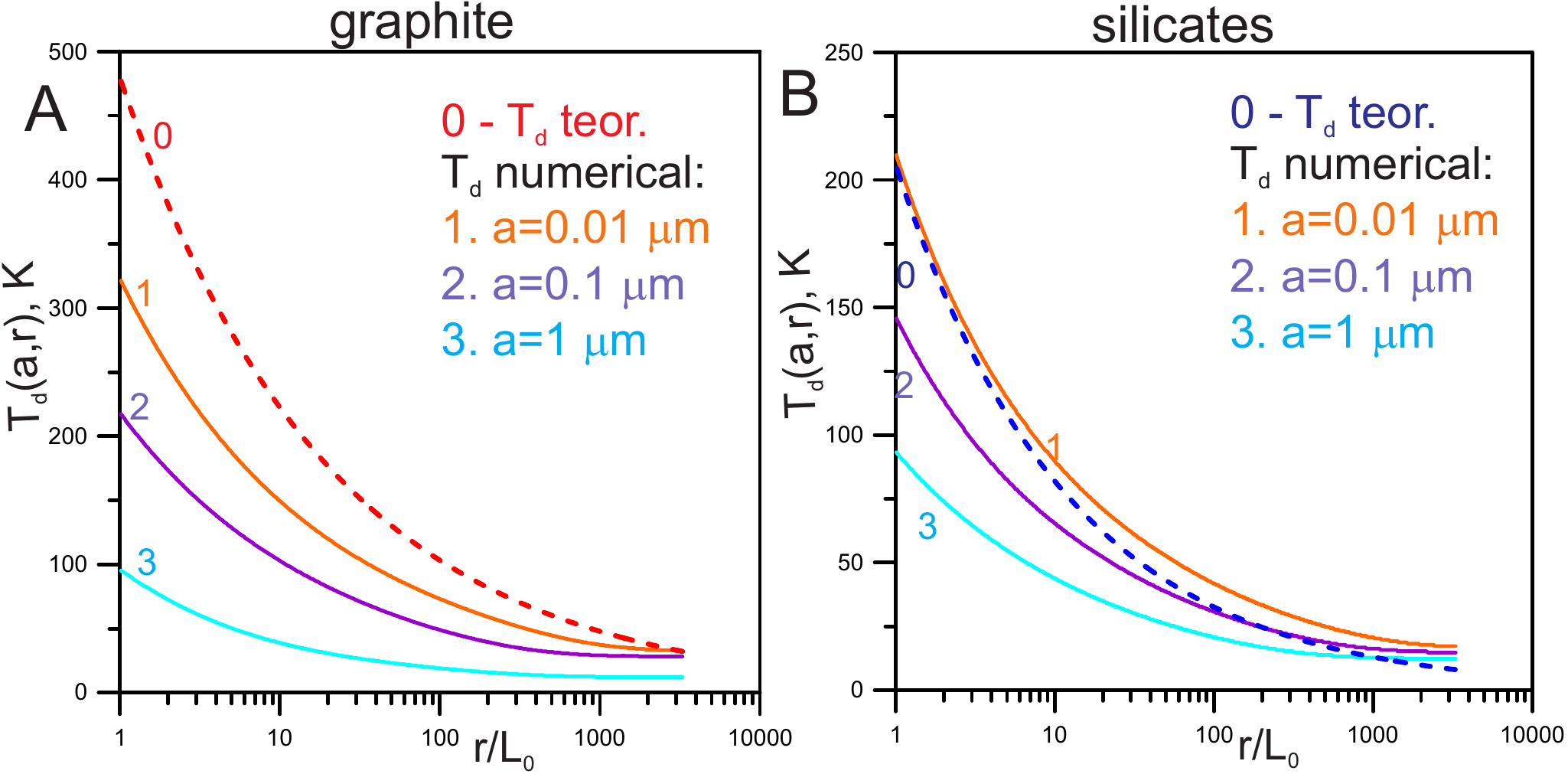}
\centering
\caption{Temperature of dust grains as a function of distance from the star for graphite (plot~A) and silicates (plot~B) calculated for the system $\sigma$~Ori~AB.
Dashed curves correspond to the analytical approach for $T_d(r)\sim r^{-\gamma}$ (see Appendix~\ref{app:Td}).
Solid curves correspond to numerical solution for different grains radius.
}
\label{fig:Td_r}
\end{figure*}

Dust grain temperature $T_d$ can be calculated either analytically in the frame of the Rayleigh scattering approach (in this case $T_d$ depend only on $r$ that is a distance from the star,
see Appendix~\ref{app:Td} for details), or numerically by solving an equation of thermal equilibrium for each grain (dust grains heat up due to
absorption of stellar and interstellar radiation and cool down due to the thermal black-body emission). Numerical solution for $T_d(a,r)$ is described by Katushkina et al. (2018).
As an example we perform calculations of the dust temperature around the system $\sigma$~Ori~AB.
Radius and effective temperature for three components of $\sigma$~Ori~AB are presented in Table~4 of Sim\'{o}n-D\'{i}az et al. (2015).
In our calculations we assume that the effective radiation spectrum of $\sigma$~Ori~AB is a sum of individual spectra of its components.
Comparison of the dust temperature obtained by two approaches for the system $\sigma$~Ori~AB is presented in Fig.~\ref{fig:Td_r}.
 In general, it is seen that
temperature for graphite is systematically larger than for silicates (as we will see below it is extremely important for intensity of
infrared emission from dust). For graphite the theoretical approach overestimates the temperature for all distances and sizes, while for silicates theoretical approach provides larger temperature at small distances and smaller temperature far away from the star
compared with the numerical one for typical sizes of grains.

We are interested in emission from the astrosphere, namely, from the region, where the dust distribution is disturbed
compared with the interstellar medium. In the undisturbed ISM the dimensionless dust number density is $\hat{n}=1$ and temperature of dust grains ($T_{d,ISM}(a)$) is determined by isotropic interstellar radiation (Hocuk et al. 2017).
In the numerical model we consider a cubic box around the star with a size of $d_{max}=3r_{max}$, where $r_{max}=4/a_{min}$, $[a_{min}]=\mu m$,
$r_{max}$ is the largest radius of the outer arc for the smallest considered dust grain. An integration of intensity
for the realistic dust distribution ($I_{star}$) is performed within the box. We also calculated an intensity that would be produced in the box if it is filled by
the undisturbed interstellar dust ($I_{ISM}$). The difference $I_{A}=I_{star}-I_{ISM}$ provides an excess of the intensity due to changes of the dust distribution
in the vicinity of the star and can be expressed by the following integral:
\begin{eqnarray}\label{intensity}
 I_{A}(\lambda_0) = 2 \cdot \int_{0}^{d_{max}} \int_{a_{min}}^{a_{max}} \pi a^2 C_{ISM} F_{ISM}(a) Q_{abs}(a, \lambda_0) \cdot \\
  \cdot [\hat{n}(\frac{r}{D^*(a)}, \theta) \cdot B_{\nu_0}(T_d(a,r)) - B_{\nu_0}(T_{d,ISM}(a))] da \, dy  \nonumber
\end{eqnarray}
Factor 2 is associated with the axisymmetry of the problem (integration over $y$ is performed in a range of $[0, d_{max}]$, while in principle we need to
integrate from $-d_{max}$ to $+d_{max}$). We also perform test calculations with two times larger integrational box and results are the same
that means that chosen value of $d_{max}$ is appropriate.

In order to calculate dimensional intensity of thermal emission one needs to specify certain stellar and interstellar parameters. Namely,
to calculate $C_{ISM}$ we assume that the protons number density in the undisturbed ISM is $n_{p,ISM}=5\, cm^{-3}$,
the gas to dust mass ratio is 100, and the
 density of dust grains material is 2.5~$g/cm^3$.

\section{Results: infrared intensity maps}

 In this section we present the results of calculations of the intensity maps for wavelength $\lambda_0=24~\mu m$.
 In general, we have four parameters of the model, which influence the results: 1) material of dust grains (graphite or silicates);
 2) parameter $\beta(a)$ calculated either analytically ($\sim 1/a$) or numerically; 3) kind of a size distribution
  of dust in the ISM (``MRN'' with $\alpha_0$=2.5 or 3.5, Mathis 1996 or Artymowicz \& Clampin, 1997) and 4) temperature of dust grains calculated either
 analytically in the Rayleigh scattering approach ($T_d(r)$, see Appendix~\ref{app:Td}) or numerically $T_d(a,r)$.

 \begin{figure*}
\includegraphics[scale=0.7]{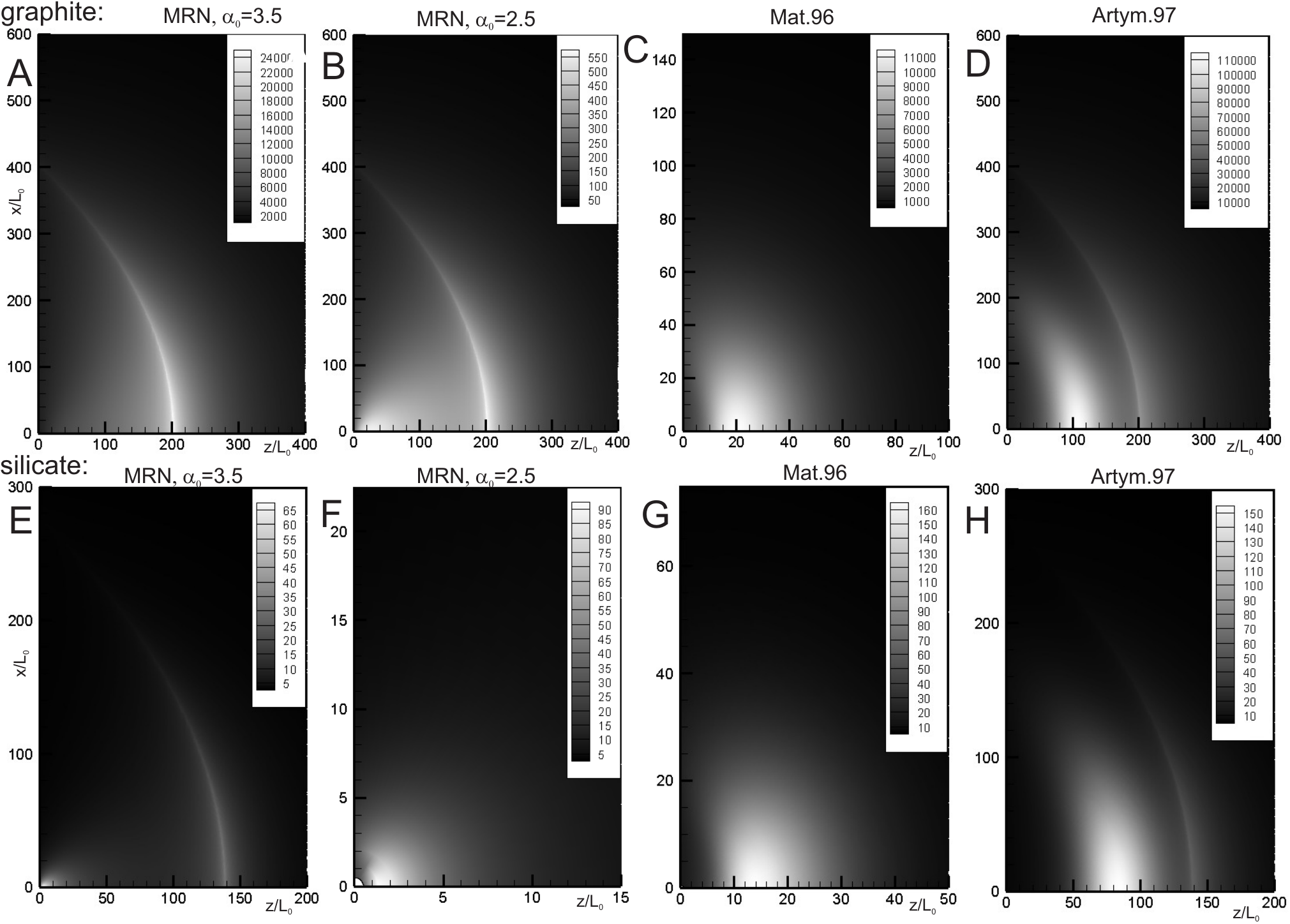}
\centering
\caption{Maps of intensity $I_{A}$ in MJy/sr obtained for graphite and silicates with four different size distributions of dust in the ISM.
 These maps are obtained with numerically calculated $\beta(a)$ and numerically calculated dust temperature $T_d(a,r)$.
}
\label{fig:intens_FISM}
\end{figure*}

Fig.~\ref{fig:intens_FISM} shows the intensity maps obtained with different dust size distributions in the ISM and with numerically calculated
$\beta(a)$ and $T_d(a,r)$.
For graphite the distinct arc is seen for all cases except Mathis (1996) size distribution.
For silicates the arc exists for MRN with $\alpha_0=3.5$ and Artymowicz \& Clampin (1997) size distributions.
Emission of dust is determined by both the dust temperature and the number density. I.e. if the dust grains within the arc with the enhanced number density
(see Fig.~\ref{fig:n_mix_bet}) are heated enough then the arc is visible at the intensity maps.
The dust temperature depends on several competitive factors. On the one hand, dust grains with the same size located closer to the star are heated more than those further from the star,
but on the other hand, only large dust grains can penetrate close to the star and they are heated less than the smaller ones at the same distance as it is seen from
Fig.~\ref{fig:Td_r}.
Therefore, for graphite in the case of MRN size distribution (plots A and B in Fig.~\ref{fig:intens_FISM}) the distinct arc is seen exactly
at the same place where the dust number density is enhanced (compare with plots A and B in Fig.~\ref{fig:n_mix_bet}), because these arcs are produced
by the smallest dust grains with the highest temperature.
In the case of Mathis (1996) size distribution the arc of enhanced number density is formed by quite large dust grains
(the largest one within all considered size distributions) with $a_{peak}$=0.1~$\mu m$ and the number density at the arc
is quite small (see Fig.~\ref{fig:n_mix_bet}~C and G). As the result, this arc is not seen at the intensity maps, because
dust grains are not heated enough (plots~C and G in Fig.~\ref{fig:intens_FISM}). However, larger dust grains, which penetrate even closer to the star, are heated there by the stellar radiation
and provide a bright maximum of emission at small distances from the star (see plots C and G in Fig.~\ref{fig:intens_FISM}).
In the intermediate case of Artymowicz \& Clampin (1997) size distribution the arcs at $z/L_0$=200 for graphite and $z/L_0$=140 for silicates
are seen at the intensity maps. These arcs are located further from the star than the arcs of enhanced number density (see plots~D and H in Fig.~\ref{fig:n_mix_bet}),
because they are formed by smaller and hotter dust grains.
Besides this, there are bright maxima of intensity a bit closer to the star that are produced by
a little bit larger dust grains, which are heated enough to be visible.
Silicates are less heated than graphite, and therefore arcs are not seen at the intensity maps in plots~F-G in Fig.~\ref{fig:intens_FISM}.
Note that due to the same reason absolute values of intensity for graphite are several order of magnitude larger than for silicates.


\begin{figure*}
\includegraphics[scale=0.7]{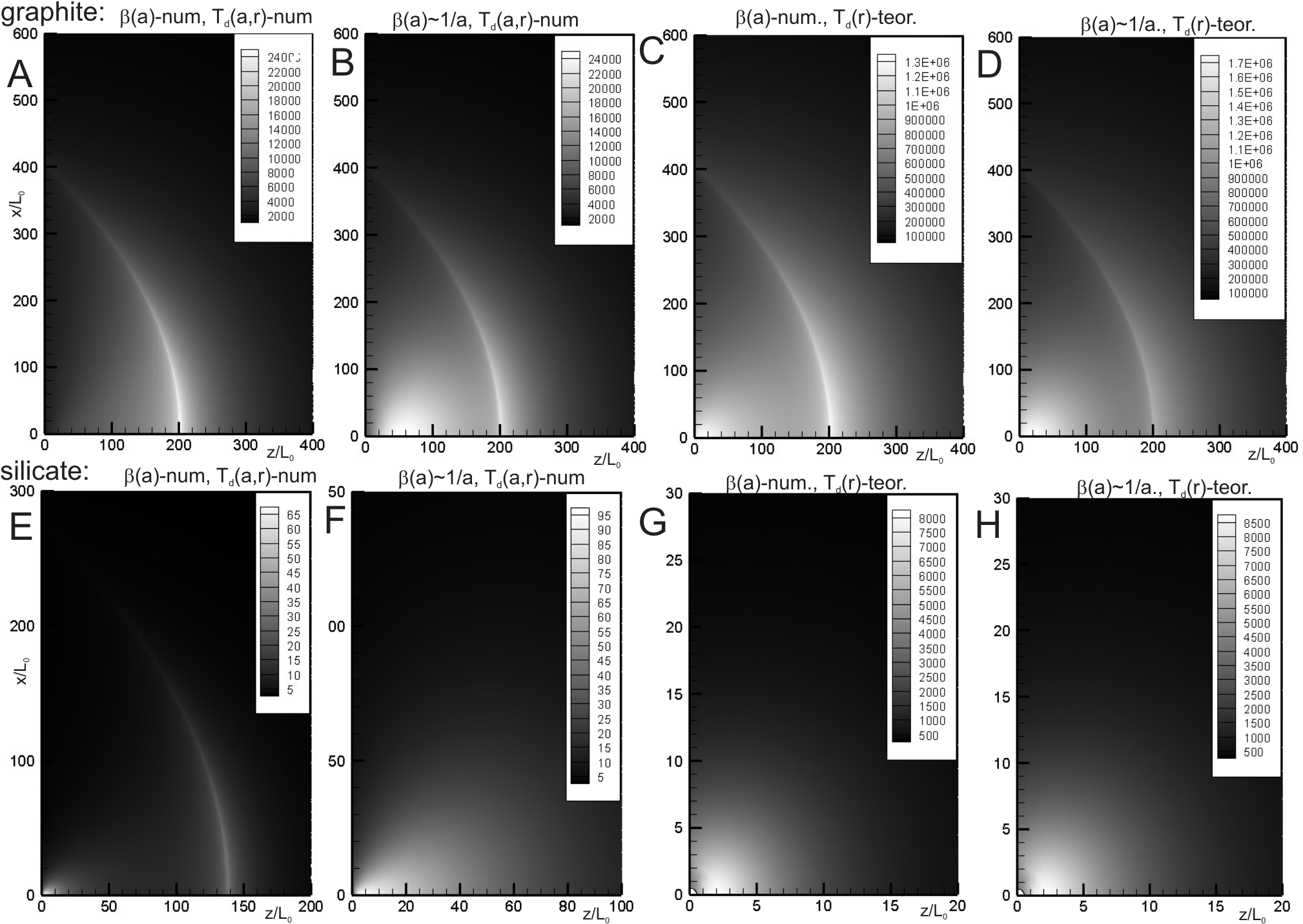}
\centering
\caption{Maps of intensity $I_{A}$ in MJy/sr obtained for graphite (plots A-D) and silicates (plots E-H) with two ways of calculations of
the dust temperature: numerical one (plots A,B,E,F) and analytical one (plots C,D,G,H) and
 two ways of calculation of $\beta(a)$: numerical one (plots A,C,E,G) and analytical $\beta(a)\sim1/a$ (plots B,D,F,H). These maps are obtained with classical MRN size distribution in the ISM with $\alpha_0=3.5$.
}
\label{fig:intens_Td}
\end{figure*}

 Fig.~\ref{fig:intens_Td} illustrates an effect of the way of calculation of the dust temperature and $\beta(a)$ on the intensity maps.
 Namely, it presents the intensity maps obtained for graphite (plots A-D) and silicates (plots E-H) with $\beta(a)$ and the dust temperature calculated
   numerically and analytically and the same MRN dust distribution in the ISM.
   In general, it is seen that the bright arc is visible only for graphite. For silicates the arc with low intensity exists for numerically calculated $\beta$ and $T_d$, while all other maps look the same with one bright maximum
   close to the star. This is explained by the fact that the dust temperature for silicates is less than for graphite (see Fig.~\ref{fig:Td_r}), and only
  dust grains close to the star emit enough infrared radiation to be visible. The visible difference for silicates (plots F-H in Fig.~\ref{fig:intens_Td}) is only in absolute values of intensity, which are about 100 times larger in the case of
   theoretical approach for $T_d(r)$ compared with the numerical one $T_d(a,r)$.
   Intensity obtained for silicates is two order of magnitude less than for graphite.
   Position of the bright thin arcs seen at the intensity maps for graphite (plots A-C in Fig.~\ref{fig:intens_Td}) correspond to the arcs in the dust number density distribution, where the density is increased (see Fig.~\ref{fig:n_mix_bet}~A and Fig.~\ref{fig:n_mix}~A).

  It is interesting that using the ``theoretical'' approach for $\beta(a)$ for graphite (plot~B in Fig.~\ref{fig:intens_Td}) instead of
   numerical one leads to
   appearing of a bright maximum close to the star at $z/L_0$=40-50.
   This is because of the following reason: for dust grains with $a\in[0.01;0.7]$~$\mu m$ the theoretical $\beta(a)$ is smaller than the numerical one (it is seen from Fig.~\ref{fig:beta_nism}~A),
   therefore more amount of grains penetrate closer to the star and provide an additional maximum of intensity.

   Using the ``theoretical'' approach for the dust temperature in the case of graphite (plot~C in Fig.~\ref{fig:intens_Td}) instead of
   the numerical one also leads to
   appearing of a very bright maximum close to the star at $z/L_0\sim 20$, and the arc at $z/L_0$=200 also exists with much larger intensity.
   This is because of the theoretical dust temperature is much larger than the numerical one close to the star (see Fig.~\ref{fig:Td_r}~A). Small distances near the star can be approached only by quite large dust grains, which are less
   heated in the case of numerically calculated dust temperature, therefore this local maximum of intensity is absent in plot~A in Fig.~\ref{fig:intens_Td}.
    In general, using theoretical Rayleigh scattering approach for calculations of the dust temperature leads to overestimation of the
   intensity of the infrared emission by 1-2 orders of magnitude.

\section{Conclusions}

In this work we consider an effect of the stellar radiation pressure on the interstellar dust distribution and intensity maps of the infrared
thermal dust emission. This situation is applicable for the stars with relatively weak stellar wind and strong radiation.
In this case the astropause and the bow shock are located close to the star while the stellar radiation swept out the dust much further from the star.
Therefore, the observed dust arcs are formed due to an interaction of the interstellar dust flow with the stellar radiation rather than with the stellar wind.

The numerical model of the dust distribution is presented. The model has several parameters such as material of dust grains, kind
of size distribution of dust in the undisturbed interstellar medium, way of calculation of the ratio of the stellar radiation to the stellar gravitation
$\beta(a)$ and the interstellar dust temperature either analytically (in the frame of simplified approaches) or numerically.

It is found that the dust number density distribution has a distinct arc with enhanced number density by 10-300~\% within the arc.
The shape of the arc is parabolic, and its location (distance from the star) depends on the assumed size distribution of dust in the ISM and
the applied $\beta(a)$ values. The arc of enhanced dust number density exists for all considered model parameters.

The intensity maps of the thermal infrared emission from dust are calculated for different sets of the model parameters.
Intensity of the infrared emission strongly depends on the dust temperature, which in turn depends on distance from the star and radius of dust grains.
Particulary, smaller grains are heated more than the larger ones (at certain distance form the star), while the large grains can penetrate close to the star, where
the strong stellar radiation may them heated up more than far away from the star. These two competing effects may reduce
an intensity of radiation from the arc with enhanced dust number density or even the absence of a visible arc.
The main results related to infrared arcs are summarized by the following:
\begin{itemize}
  \item intensity of infrared thermal radiation
for graphite is several order of magnitude larger than for silicates, because silicates are less heated due to less absorption coefficients;
  \item distinct bright arcs at the infrared intensity maps (24~$\mu m$) are obtained mostly for graphite, while for silicates there is only a
  maximum of intensity close to the star instead of arcuate structure;
  \item when the arc is visible, it has a parabolic shape with $x_0=2\cdot z_0$ ($x_0$ and $z_0$ are coordinates where the arc intersects axis X and Z);
  \item intensity of thermal emission strongly (up to 200 times) depends on the assumed size distribution of dust in the ISM; in the case
  of graphite the distribution of Artymowicz \& Clampin (1997) provides the largest intensity;
  \item using theoretical Rayleigh scattering approach for calculations of the dust temperature $T_d(r) \sim 1/r^{\gamma}$ leads to overestimation of the
   intensity of the infrared emission by two order of magnitude.
\end{itemize}

We conclude that by analyzing the observed images of the astrospheres one needs to be careful and choose appropriate parameters of the model.
The composition and size distribution of the interstellar dust are not known precisely for many certain parts of the ISM.
Since the visible intensity is sensitive to these parameters than it is possible to use the observations to obtain constrains
for the interstellar dust distribution around certain stars. Here we provide a kind of ``algorithm'' that can be used for analysis of the
observational image of the astrosphere if the distinct arc is visible:
\begin{itemize}
  \item to estimate dimensionless parameters in the dust motion equation one needs to know (or estimate somehow) dimensional parameters of the star
  ($M_s$, $L_s$, $\dot{M}$, $v_{\infty}$) as well as parameters of the interstellar plasma ($V_{ISM}$, $n_{p,ISM}$, $T_{p,ISM}$, $B_{p,ISM}$) and dust grains ($U_d$, $\rho_d$).
  After that it is possible to calculate the distances $L^*$, $L_1$, $L_2$, $L_3$ for dust grains with typical radius $a=0.01-1$~$\mu m$.
  Comparison of three dimensionless parameters $L_1/L^*$, $L^*/L_2$ and $L^*/L_3$ with each other shows which force dominates outside of the astropause;
  \item if the central radiation pressure force dominates for all typical dust grains sizes then it is possible to describe the astrosphere by using of the
  ``cold model'' considered here;
  \item in the frame of the model it is possible to calculate maps of the infrared emission from dust under some assumptions about dust
  material and size distribution in the ISM;
  \item comparison of the model results with the observations may be used to determine appropriate parameters of the model that gives
  an agreement between the model and data.
\end{itemize}

Note, that it is necessary to take into account an angle between direction of the star's motion
  through the ISM and our line of sight. I.e. we could observe the astrosphere from flanks or from the nose and it should be taken into account in the model.
  Effect of geometrical projection of the arc at different planes is discussed in Appendix of Gratier et al. (2014).

In future works we will consider the cases when two of three forces play a significant role and should be considered simultaneously.


\section*{Acknowledgments}

\clearpage

\appendix

\section{Analytical integration of the dust number density for mixture of different grains in the frame of the cold model}
\label{app_A}
In this section we show mathematically how to calculate an integral of the dust number density for mixture of dust grains with different radii
under the following assumptions:
\begin{itemize}
  \item the classical MRN size distribution of dust in the ISM;
  \item $\beta(a)\sim1/a$ (an approach of geometrical optics);
  \item $\beta(a)\gg1$;
\end{itemize}
The character distance is chosen as $L_0\approx GM_s\beta(a_0)/V_{ISM}^2$ for $a_0=1 \mu m$.

As it follows from Section~3.2, the dust number density for mixture of different grains is
\begin{equation}
 n_{d,mix}(\hat{r},\theta) = \int_{a_{min}}^{a_{max}} \hat{n}_d(\hat{r}\cdot \frac{a}{a_0}, \theta)C_{ISM}a^{-\alpha_0}da
\end{equation}
Let us note that $a/a_0=a$ for $a_0=1 \mu m$ and $[a]=\mu m$. Hereafter we omit $a_0$ where it is possible, and assume that the grain's radius
$a$ is expressed in $\mu m$ by default.
Let's make several mathematical transformations:
\[
 \hat{n}_d(\hat{r} \cdot a, \theta)=\frac{1+A}{2\sqrt{A}},
\]
where
\[
  A(\hat{r} \cdot a, \theta)=1-\frac{4}{\hat{r}(1+\cos(\theta))} \cdot \frac{1}{a}=
 1-\frac{B(\hat{r},\theta)}{a},
\]
and $B(\hat{r},\theta)=4/(\hat{r}(1+\cos(\theta))$.
Hence
\[
 \hat{n}_d(\hat{r} \cdot a, \theta) = \frac{2a-B}{2\sqrt{a(a-B)}}, \text{and}
\]
\[
  n_{d,mix}(\hat{r},\theta) = C_{ISM} \, \int_{a_{min}}^{a_{max}} \frac{2a-B}{2\sqrt{a(a-B)}}\cdot a^{-\alpha_0} da.
\]

For each radius of dust grain there is a cavity near the star where these grains can not penetrate.
The size of the cavity depends on the grain's radius, namely, bigger grains penetrate closer to the star. The boundary of the cavity
formed by grains with radius $a$ is the arc with the following equation (in considered dimensionless units):
\[
  \hat{r}(\theta)=\frac{4}{(1+\cos(\theta))\cdot a}.
\]
Let us consider the inner arc formed by the largest dust grains, $\hat{r}_1(\theta)$, and the outer arc formed by the smallest dust grains,
$\hat{r}_2(\theta)$.

If $\hat{r}(\theta)<\hat{r}_1(\theta)$ for a given $\theta$ then $n_{d,mix}(\hat{r},\theta) = 0$, because no dust grains can penetrate into the smallest cavity.

If $\hat{r}(\theta)\in[\hat{r}_1(\theta), \hat{r}_2(\theta)]$ then there is a grain's radius $a^*\in[a_{min}, a_{max}]$ such as:
\[
  \hat{r}(\theta)=\frac{4}{(1+\cos(\theta))\cdot a^*},
\]
i.e. $B(\hat{r},\theta)=a^*$. This means that at this point $(\hat{r},\theta)$ there are dust grains with radius $a\in[a^*,a_{max}]$, because
smaller grains are swapped out further from the star. Therefore,
\[
  n_{d,mix}(\hat{r},\theta) = C_{ISM} \, \int_{a^*}^{a_{max}} \frac{2a-a^*}{2\sqrt{a(a-a^*)}}\cdot a^{-\alpha_0} da.
\]

If $\hat{r}>\hat{r}_2$ then dust grains with all radii exist at this point and integration should be performed from $a_{min}$ to $a_{max}$.

The considered indefinite integrals can be calculated analytically for $\alpha_0=3.5$:
\begin{eqnarray}
 \int\frac{2a-B}{2\sqrt{a(a-B)}}\cdot a^{-3.5} da = \nonumber \\
 = \frac{7y(3y^2+5B)(y^2+B)-8yB^2}{48B^2(y^2+B)^3}+\frac{21}{48 B^2\sqrt{B}}\arctan\frac{y}{\sqrt{B}}, \nonumber
\end{eqnarray}
and for $\alpha_0=2.5$:
\begin{eqnarray}
 \int \frac{2a-B}{2\sqrt{a(a-B)}}\cdot a^{-2.5} da = \nonumber \\
 =  \frac{y(5y^2+3B)}{8B(y^2+B)^2}+\frac{5}{8B\sqrt{B}}\arctan \frac{y}{\sqrt{B}}, \nonumber
\end{eqnarray}
where $y=\sqrt{a-B}$.

Therefore, the definite integrals with any bounds can be easily calculated.
This means that although the number  density distribution of dust grains with each certain radius is infinite
at the corresponding arc, but this singularity is integrable and the total number density distribution of mixture
of different grains does not contain a singularity.

\section{Analytical approach for the dust grains temperature}
\label{app:Td}

The energy balance for a single dust grain can be expressed by the following equation (see e.g. Hocuk et al. 2017):
\begin{eqnarray}\label{eq:Td}
4\pi\,\int_0^{\infty}Q_{abs}(a,\nu)B_{\nu}(T_{d})\,d\nu=\int_0^{\infty} Q_{abs}(a,\nu)F^*_{\nu}(r,T_*)\,d\nu \nonumber,
\end{eqnarray}
where $F^*_{\nu}=\pi\, (R_*/r)^2 B_{\nu}(T_*)$ is the stellar radiation spectrum at distance $r$ from the star ($B_{\nu}$ is a Plank function),
$R_*$ is the stellar radius, $T_*$ is the stellar effective temperature (in our calculations we took these parameters for $\sigma$~Ori~AB system).
Here we neglect an effect of the interstellar radiation and assume that the dust grains heats up only by stellar radiation and cool down by thermal emission.

It is convenient to consider the Plank mean opacity:
\begin{eqnarray*}
 k_p(T) = \frac{\int_0^{\infty}Q_{abs}(a,\nu)B_{\nu}(T)d\nu}{\int_0^{\infty}B_{\nu}(T)d\nu}= \\ \nonumber
 =\left(\frac{\sigma_{SB}\, T^4}{\pi}\right)^{-1}\int_0^{\infty}Q_{abs}(a,\nu)B_{\nu}(T)d\nu,  \nonumber
\end{eqnarray*}
$\sigma_{SB}$ is the Stefan-Boltzmann constant.
Thus, an equation~(\ref{eq:Td}) can be rewritten as:
\[
 4\pi \cdot k_p(T_d)\cdot T_d^4 = \pi \cdot \left(\frac{R_*}{r}\right)^2 \cdot k_p(T_*)\cdot T_*^4,
\]
hence
\begin{equation}\label{eq:Td_analyt}
 T_d=T_* \cdot \left(\frac{R_*}{2r}\right)^{1/2}\left(\frac{k_p(T_*)}{k_p(T_d)}\right)^{1/4}.
\end{equation}
If one knows a function $k_p(T)$ then this equation can be solved numerically.

However, it is possible to solve this equation analytically in the Rayleigh scattering approach (when wavelength $\lambda$ is much larger than
the grain's size $2\pi a$). In this case $Q_{abs}(a,\nu)\sim (a \nu)^\kappa$, where $\kappa=1$ for astronomical silicates and $\kappa=2$ for graphite.
Therefore equation~(\ref{eq:Td_analyt}) can be integrated and temperature $T_d$ can be expressed as:
\begin{equation}
 T_d(r)=T_*\cdot\left(\frac{R_*}{2r}\right)^\gamma,
\end{equation}
where $\gamma=2/5$ for silicates and $\gamma=1/3$ for graphite. It is seen that under the Rayleigh scattering approach
dust grains temperature does not depend on the grain's radius.
As it is shown in Fig.~7~A in Katushkina et al. (2018) this is not true for typical interstellar dust grains. However, we consider this approximation here, because sometimes it is used in the literature (Decin et al., 2006, Gvaramadze et al., 2019).

\label{lastpage}

\end{document}